\documentclass[fleqn,usenatbib,twocolumn]{mnras}

\usepackage{txfonts}

\usepackage[T1]{fontenc}

\DeclareRobustCommand{\VAN}[3]{#2}
\let\VANthebibliography\thebibliography
\def\thebibliography{\DeclareRobustCommand{\VAN}[3]{##3}\VANthebibliography}

\usepackage{graphicx}	
\let\iint\relax

\usepackage{amsmath}	

\usepackage{afterpage}

\usepackage{xcolor}

\newcommand{\I}[0]{\text{i}}
\newcommand{\dd}[1]{\text{d}{#1}}

\DeclareMathOperator{\sinc}{sinc}

\defcitealias{Huppenkothen2018}{H18}
\defcitealias{Bachetti2018}{B18}

\title[Statistical properties of the cross spectrum]{The statistical properties of the cross spectrum}

\author[E. Nathan et al.]{
Edward J. R. Nathan,$^{1}$\thanks{E-mail: edward.nathan@nasa.gov }
Adam Ingram,$^{2}$
Daniela Huppenkothen,$^{3,4}$
Matteo Bachetti$^{5}$, and Javier A. Garc\'ia$^{6,1}$
\\
$^1$NASA Postdoctoral Program Fellow, X-ray Astrophysics Laboratory, NASA Goddard Space Flight Center, Greenbelt, MD 20771, USA\\
$^2$School of Mathematics, Statistics, and Physics, Newcastle University, Newcastle upon Tyne, NE1 7RU, UK\\
$^3$Anton Pannekoek Institute for Astronomy, University of Amsterdam, Science Park 904, 1098XH Amsterdam, the Netherlands \\
$^4$SRON Netherlands Institute for Space Research, Niels Bohrlaan 4, 2333 CA Leiden, the Netherlands\\
$^5$INAF-Osservatorio Astronomico di Cagliari, via della Scienza 5, I-09047, Selargius (CA), Italy\\
$^6$X-ray Astrophysics Laboratory, NASA Goddard Space Flight Center, Greenbelt, MD 20771, USA
}

\date{Accepted XXX. Received YYY; in original form ZZZ}

\pubyear{2015}

\begin{document}
\label{firstpage}
\pagerange{\pageref{firstpage}--\pageref{lastpage}}
\maketitle

\begin{abstract}
The cross spectrum encodes the correlated variability between two time signals. In recent years, the cross spectrum has been used to study astronomical sources, particularly in the field of X-ray timing. In the literature, it has been common to either simultaneously fit the real and imaginary components of the cross spectrum, or fit the phase and magnitude. Until now, a full discussion of the statistical distribution of the cross spectrum has been missing from the astronomical literature.
In this paper, we present a derivation of the full statistical distribution of a cross spectrum between two time series, showing that it follows an asymmetric Laplace distribution. We further provide the probability distribution function for a cross spectrum random variable, along with the marginal distributions for many quantities.  We also relate the cross spectrum to the power spectra of the constituent time series.
This work will enable the cross spectrum to be used more accurately as a probe of physical processes such as accretion onto black holes and neutron stars.
\end{abstract}





\section{Introduction}

Fourier analysis is a powerful toolbox which has been employed in many areas of astronomy to study variability.  In many cases the processes behind the observed variability can be considered both stochastic, and (on the timescales of observations) stationary.

In the context of X-ray astronomy, spectral timing studies have linked the observed variability to the spectral information.  A common tool used in these studies is the cross spectrum, a complex quantity that correlates variability in two time series, typically light curves of different X-ray energy bands \citep[e.g.][]{Uttley2014, Rapisarda2016, Mastroserio2018, Nathan2022, Mendez2024}.  As a complex quantity, the phase of the cross spectrum gives the phase lag between the two signals, while the magnitude is related to the correlated power of the signals \citep[which can be normalised to their coherence;][]{Vaughan1997}.
The real part of the cross spectrum is also known as the cospectrum, while the imaginary part is the quadrature spectrum.  When two light curves are expected to be intrinsically identical (such as light curves recorded by two identical detectors), the cospectrum can be used to estimate the power spectrum when extrinsic noise is removed \citep[][hereafter \citetalias{Huppenkothen2018}]{Huppenkothen2018}.

The statistics of the power spectrum have been long known, and used within astronomy: a power spectrum random variable follows a scaled $\chi^2_2$ distribution \citep[e.g.][]{vanderKlis1989}.  However, a full description of the statistics of the cross spectrum is missing from the astronomy literature. \citet{Bendat2010} presented the expectations, error formulae, and bias terms when estimating many quantities related to the cross spectra; \citet{Epitropakis2016} discussed the statistical properties and methods of estimating time-lags with the cross spectrum. \citetalias{Huppenkothen2018} derived the statistical distribution of the cospectrum in the case where the two time series are uncorrelated; the Laplace distribution they found enables testing against a null hypothesis of cospectrum with 0 correlated variability.  Finally, \citet{Ingram2019} introduced corrections to the \citet{Bendat2010} formulae for the errors on the cross spectrum when a common reference band is used for multiple cross spectra, which is not a situation we will cover in this paper.

In Section~\ref{sec:setup} we outline the problem we are solving, and  introduce the notation we use. We also demonstrate the statistical distribution using simulations.  In Section~\ref{sec:quad_forms} we describe the cross and power spectra as quadratic forms of normal random variables, and use this to show the expectation, variance, and covariances of the spectra.  In Section~\ref{sec:crossspectrum_dist} we are able to derive the distribution of the cross spectrum, and provide the probability density function in Eq.~\ref{eqn:PDF}. In Section~\ref{sec:lowNav} we show what happens to the distributions when average quantities are used. Finally, in Section~\ref{sec:discus} we discuss the results in a common context of instruments with identical independent detectors, and provide a case study with a real observation.

\section{Problem Set-up and Notation}
\label{sec:setup}

Consider two stationary, evenly-sampled time series $\mathbf{\tilde{x}} = \{\tilde{x}_k\}_{k=1}^{K}$ and $\mathbf{\tilde{y}} = \{\tilde{y}_k\}_{k=1}^{K}$ with $K$ samples and no gaps, measured at times $\mathbf{t} = \{t_k \}_{k=0}^{K}$. For our purposes, we will generally assume that the observed time series $\mathbf{\tilde{x}}\equiv \tilde{x}(t_k)$ and $\mathbf{\tilde{y}}\equiv \tilde{y}(t_k)$ consist of a linear combination of an astrophysically relevant signal component $\mathbf{\tilde{x}}_s$ and $\mathbf{\tilde{y}}_s$ which we assume to be a stationary stochastic process, and a component of measurement noise, $\mathbf{\tilde{x}}_n$ and $\mathbf{\tilde{y}}_s$.
For now, we make the assumption that these components can all be represented by independent, normal random variables, and in general we denote random variables with a tilde, e.g. $\tilde\bullet$. Through our assumption of stationarity, the underlying means and variances of the components do not change throughout the duration of the time series.
We will be able to relax the assumption that these random variables are normally distributed later, however we will keep the assumption that the process is stationary.

We define the Discrete Fourier Transform (DFT) of a time series $\tilde{\mathcal{F}}_x(\nu_j) \equiv \mathcal{F}\big[\mathbf{\tilde{x}}\big]$, such that the time series can be written via the inverse DFT
\begin{equation}
    \tilde{\mathcal{F}}_x(\nu_j) = \sum_k \tilde{x}(t_k) \exp{\left(\I{}2\pi \nu_j t_k\right)}\,,
\end{equation}
where the $\nu_j=j/T$ is the $j^\text{th}$ Fourier frequency, and we can similarly write $\tilde{\mathcal{F}}_y(\nu_j) \equiv \mathcal{F}\big[\mathbf{\tilde{y}}\big]$.  As the Fourier terms are comprised of the sum of normal random variables, they themselves are normally distributed, albeit complex values.  We therefore can split the Fourier terms into real and imaginary parts, giving
\begin{equation}
    \begin{split}
        \tilde{\mathcal{F}}_x(\nu_j) &= \tilde{A}_{xj} + i \tilde{B}_{xj} \, \text{, and} \\ 
        \tilde{\mathcal{F}}_y(\nu_j) &= \tilde{A}_{yj} + i \tilde{B}_{yj} \,,
    \end{split}
\end{equation}

\noindent where $\I{} = \sqrt{-1}$, and $\tilde{A}_{xj}$, $\tilde{A}_{yj}$, describe the real components, and $\tilde{B}_{xj}$, $\tilde{B}_{yj}$ the imaginary components of the Fourier amplitudes at a given frequency $\nu_j$, respectively. We restrict $\nu_j$ to lie between $\nu_0 = \frac{1}{T}$, where $T$ is the total length of the time series, and the Nyquist frequency, $\nu_{K/2} = \frac{1}{2\Delta t}$ for a time resolution $\Delta t$. 

For simplicity, we have assumed that the data points $\tilde{x}_k$ and $\tilde{y}_k$ are stationary, and normally distributed around the underlying true process. 
This assumption allows us to also model the Fourier amplitudes as random variables following a normal distribution\footnote{We can relax the assumption that the time series $\mathbf{\tilde{x}}$ and $\mathbf{\tilde{y}}$ are normally distributed, provided that there are a sufficient number of data points used in the Fourier transform; the Central Limit Theorem implies that the Fourier amplitudes will be approximately Gaussian for a wide range of noise properties e.g. \citetalias{Huppenkothen2018}{}.}.

The power spectra of the two time series are the familiar $\tilde{G}_{xx,j} \equiv |\tilde{\mathcal{F}}_x(\nu_j)|^2$ and $\tilde{G}_{yy,j} \equiv |\tilde{\mathcal{F}}_y(\nu_j)|^2$.  In a similar way, the complex cross spectrum for the time series $\mathbf{x}$ and $\mathbf{y}$ is defined by
\begin{equation}
    \begin{split}
        \tilde{G}_{xy,j} & \equiv \tilde{\mathcal{F}}_x(\nu_j) \tilde{\mathcal{F}}_y^*(\nu_j) \\ 
        & = (\tilde{A}_{xj} + i \tilde{B}_{xj}) (\tilde{A}_{yj} - i \tilde{B}_{yj})  \\ 
        & = (\tilde{A}_{xj}\tilde{A}_{yj} + \tilde{B}_{xj}\tilde{B}_{yj}) +   i (\tilde{A}_{yj}\tilde{B}_{xj} - \tilde{A}_{xj}\tilde{B}_{yj})  \,.
    \label{eqn:crossspectrum}
    \end{split}
\end{equation}

We define our data series by considering the \emph{correlated} and \emph{uncorrelated} parts of the signal.  
We set $\tilde{s}_k$ as the component of the series $\tilde{x}_k$ which is perfectly linearly correlated with series $\tilde{y}_k$.  
We do not consider the case where there is a non-linear correlation of the time series, which may be detectable in higher order Fourier analysis (such as the bispectrum, \citealt{Mendel1991}).
This linearly correlated component is therefore present in series $\tilde{y}_k$ in the form $h_k * \tilde{s}_k$, where $h_k$ is the \emph{impulse-response function} that is convolved with $\tilde{s}_k$. 
This therefore leaves an uncorrelated part of each series which we denote as $\tilde{u}_{xk}$ and $\tilde{u}_{yk}$.  These terms not only include the extrinsic noise (e.g., photon counting noise) $\tilde{x}_{nk}$ and $\tilde{y}_{nk}$, but also any other intrinsic signal which happens to not be shared by the two series.  Therefore our series are $\tilde{x}_k=\tilde{s}_k+\tilde{u}_{xk}$ and $\tilde{y}_k=h_k * \tilde{s}_k + \tilde{u}_{yk}$.  
We can then write the linear decomposition of the Fourier transform of each $\tilde{x}_k$ and $\tilde{y}_k$ as
\begin{equation}
    \begin{split}
        \tilde{\mathcal{F}}_x(\nu_j) &=& \tilde{s}_j + \tilde{u}_{xj} & \\
        \tilde{\mathcal{F}}_y(\nu_j) &=& \tilde{s}_{yj} + \tilde{u}_{yj} & \equiv H_j \tilde{s}_{j} + \tilde{u}_{yj} \, ,
    \end{split}
\label{eqn:crosspec_fourier_comps}
\end{equation}
where $H_j=H_{rj} + \I{} H_{ij}$ is the transfer function: the complex but \emph{deterministic} Fourier transform of the impulse-response function.  We can then write the cross spectrum as
\begin{equation}
     \tilde{G}_{xy,j} = H_j^* \tilde{s}^{}_j\tilde{s}_j^* + \tilde{s}^{}_j\tilde{u}^{*}_{yj} + H_j^*\tilde{u}^{}_{xj}\tilde{s}_j^* + \tilde{u}^{}_{xj}\tilde{u}^*_{yj} .
    \label{eqn:rewritten_G}
\end{equation}

\subsection{Stochastic processes in the Fourier domain}

The assumptions made in the previous section mean we can consider the realisation of the Fourier terms as a draw from independent normal random variables.
The means of these distributions are 0 as we only consider stochastic processes, and due to the properties of the Fourier transform there is no covariance between the real and imaginary components.  Therefore, the variables are drawn from the distributions
\begin{equation}
    \begin{split}
        \tilde{s}_j &= \tilde{A}_{sj}+\I{} \tilde{B}_{sj}\text{, where } \tilde{A}_{js},~\tilde{B}_{js}\sim \mathcal{N}\left[0, \sigma^2_{sj}\right] \\
        \tilde{u}_{xj} &= \tilde{A}_{u_xj}+\I{} \tilde{B}_{u_xj}\text{, where } \tilde{A}_{u_xs},~\tilde{B}_{u_xs}\sim \mathcal{N}\left[0, \sigma^2_{u_xj}\right] \\
        \tilde{u}_{yj} &= \tilde{A}_{u_yj}+\I{} \tilde{B}_{u_yj}\text{, where } \tilde{A}_{u_ys},~\tilde{B}_{u_ys}\sim \mathcal{N}\left[0, \sigma^2_{u_yj}\right]\,.
    \end{split}
    \label{eqn:normal_variables}
\end{equation}

We can relate the variance of these normal distributions to the power contained within each component as e.g. $\sigma^2_s = \frac{1}{2} P_s $.

\subsection{The power and cross spectra}

\begin{figure}
    \centering
    \includegraphics[scale=0.9]{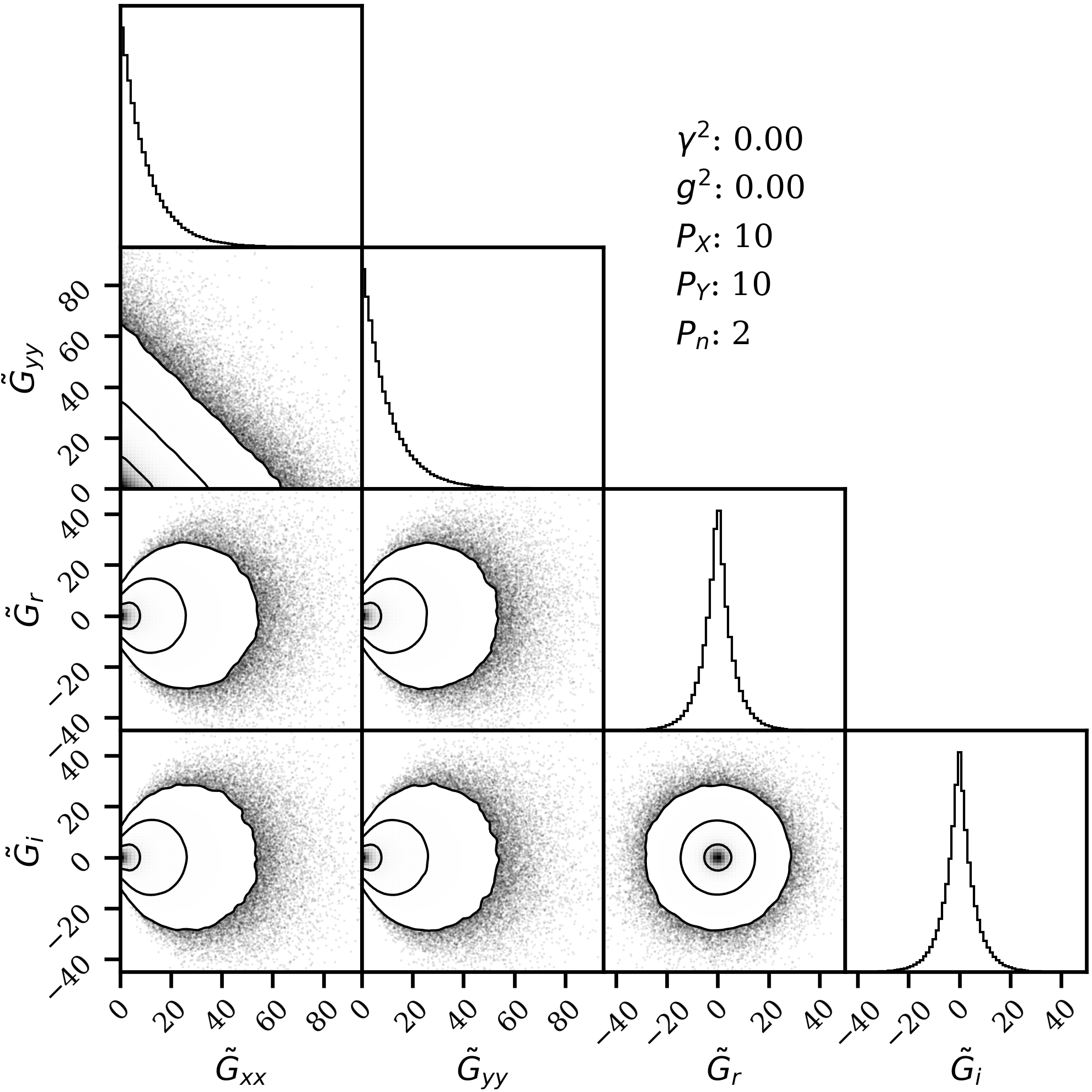}
    \includegraphics[scale=0.9]{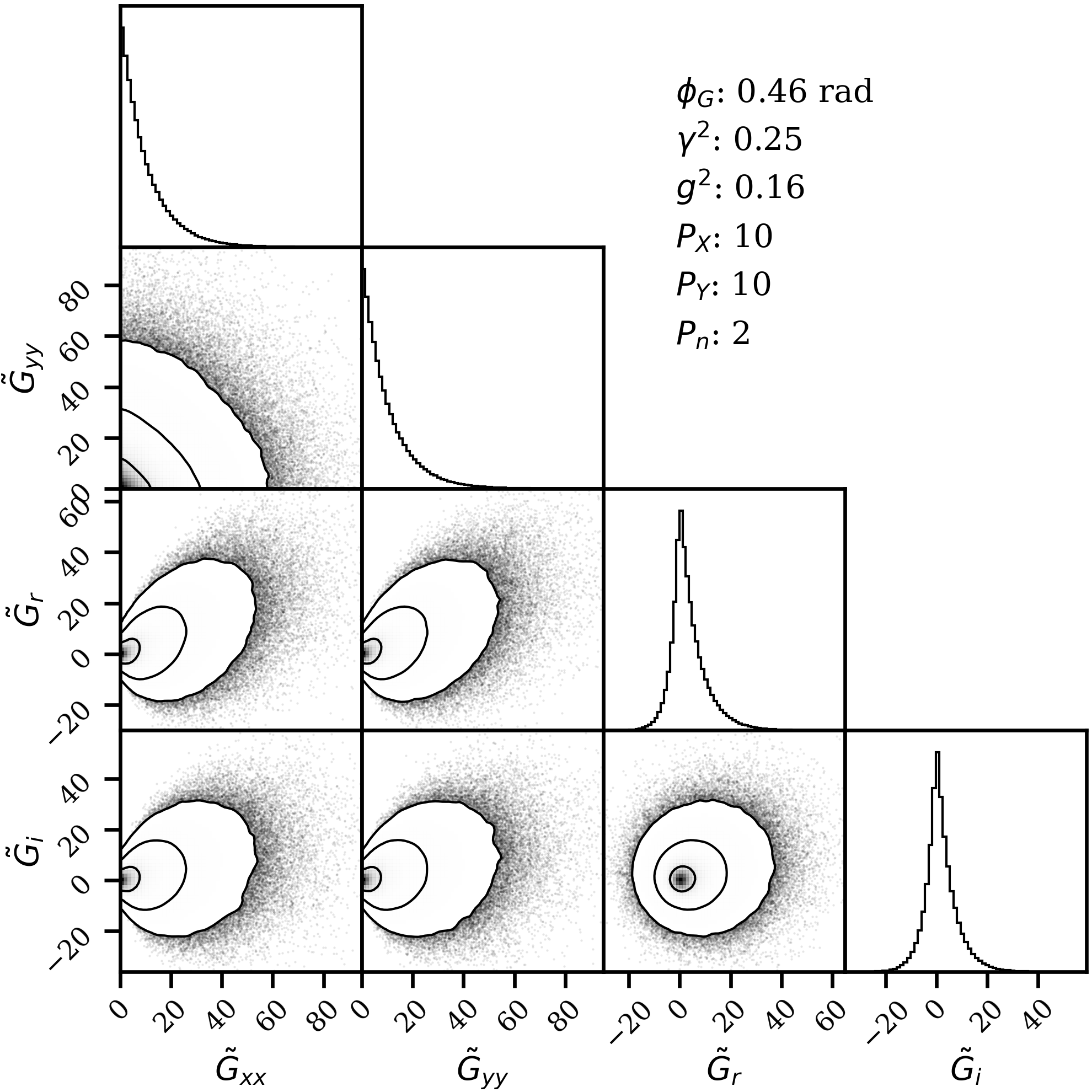}
    \includegraphics[scale=0.9]{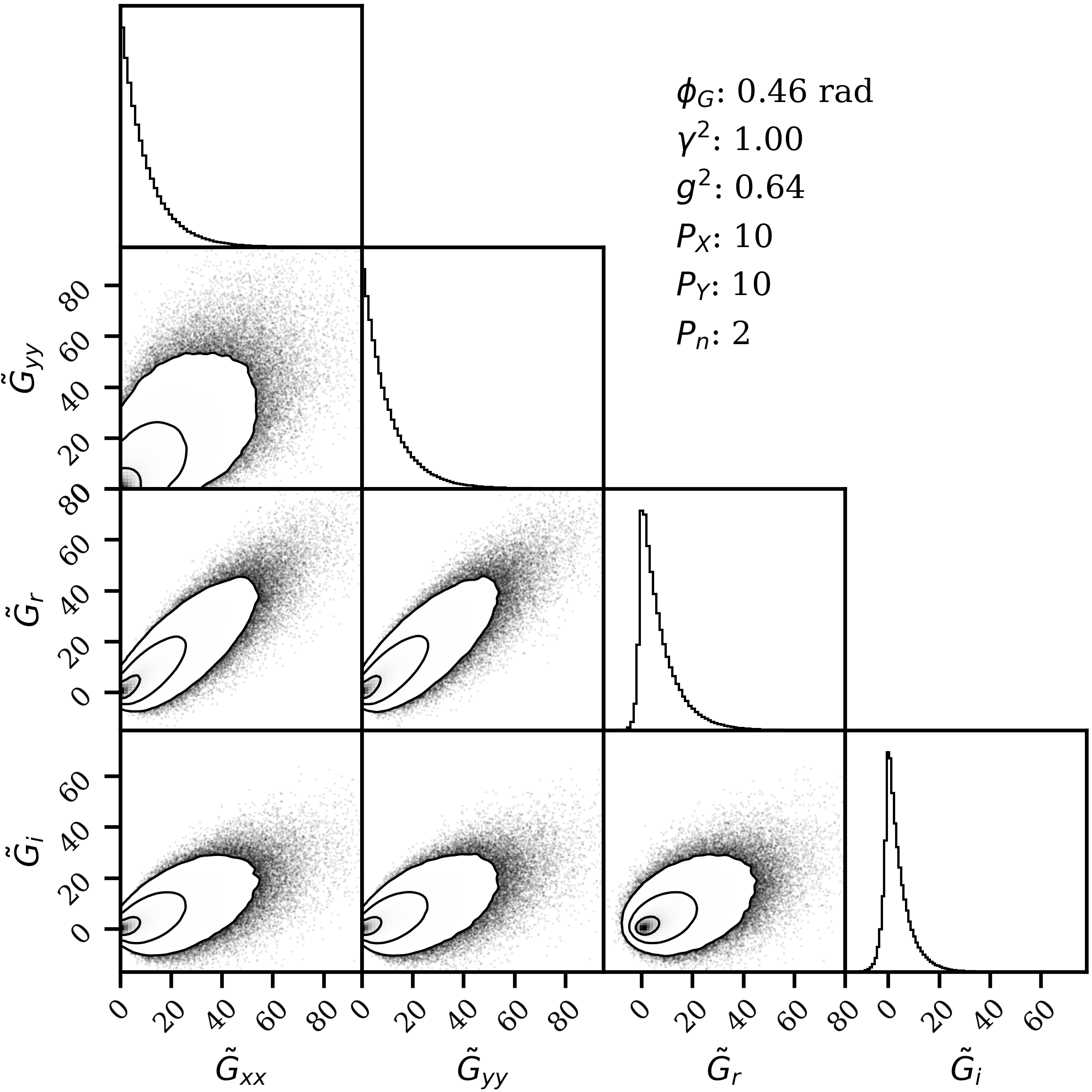}
    \caption{We show simulations of 1,000,000 realisations of a cross spectrum bin, here in an unaveraged ($N=1$) case. The simulations were performed as described in Appendix~\ref{app:simulations}. The power in the signals is $P_X=P_Y=10$, which includes a noise of $P_{n_x}=P_{n_y}=2$ (consistent with Poisson noise in Leahy normalisation). We show for three values of the intrinsic coherence ($\gamma^2=0,0.25,1$, giving the raw coherences $g^2=0,0.16,0.64$), and use a phase difference of $\phi_G=0.46$~radians.
    }
    \label{fig:PxPyGrGi_coh_sims}
\end{figure}
\begin{figure}
    \centering
    \includegraphics[scale=0.9]{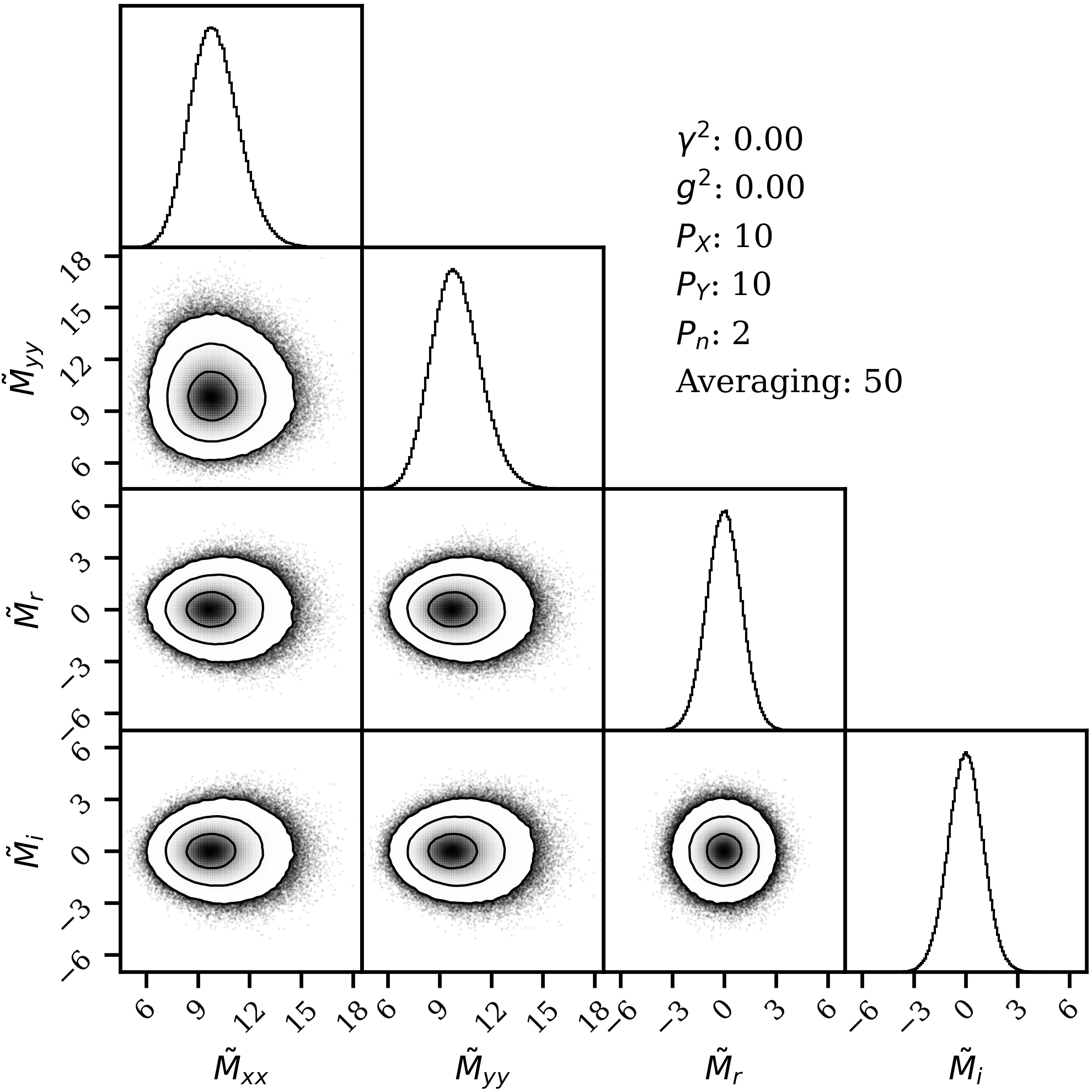}
    \includegraphics[scale=0.9]{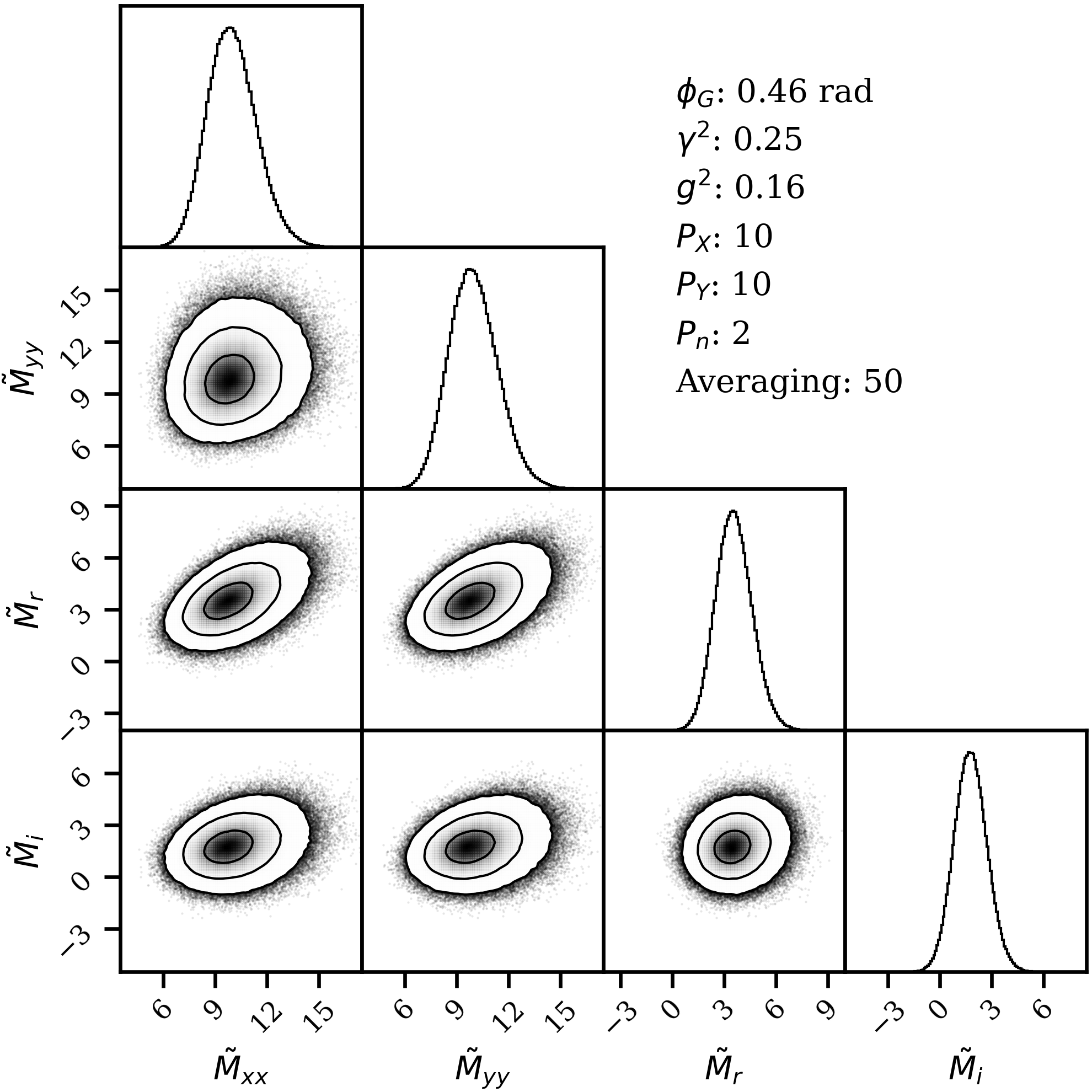}
    \includegraphics[scale=0.9]{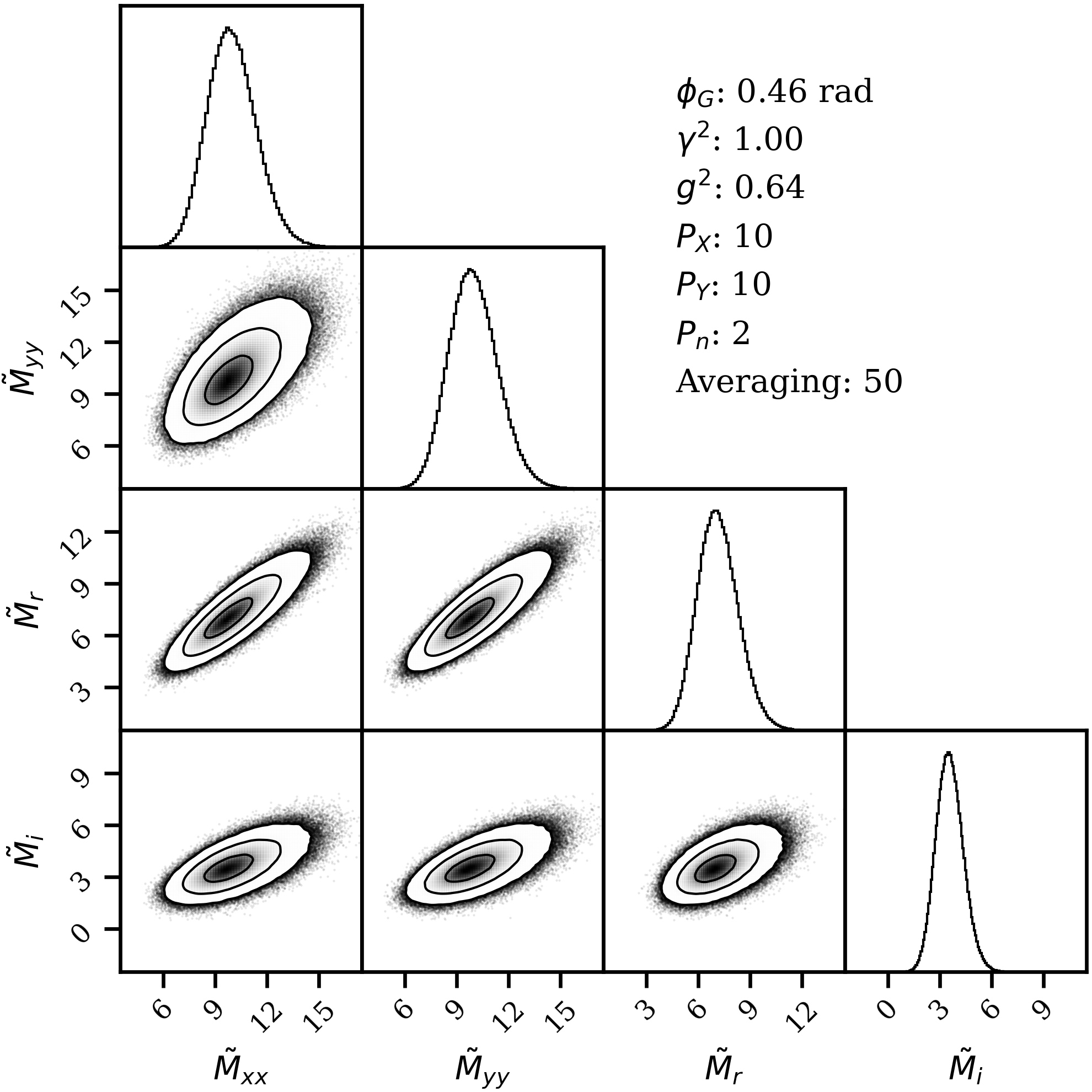}
    \caption{Similar simulations as in Fig.~\ref{fig:PxPyGrGi_coh_sims}, however this time we show the distribution of averaging of $N=50$ realisations of the cross spectrum.}
    \label{fig:PxPyGrGi_coh_ave_sims}
\end{figure}

\subsubsection{Power spectrum random variables}
The power spectrum has been written in as a quadratic form normal variables before \citep[e.g.][]{Groth1975}. However, for completeness we briefly consider the power spectral random variables 
\begin{equation}
    \begin{split}
        \tilde{G}_{xx} &= \tilde{A}_x^2 + \tilde{B}_x^2 \, \text{ and}\\ 
        \tilde{G}_{yy} &= \tilde{A}_y^2 + \tilde{B}_y^2 \, , \\ 
    \end{split}
\end{equation}
where for readability we have neglected the frequency index $j$, which applies to every term. We will continue to neglect the frequency index hereafter. Note that we distinguish between the underlying power in the signal e.g. $P_X$ with the power spectrum itself $G_{xx}$, and a random realisation of the power spectrum $\tilde{G}_{xx}$, such that we have the expectation $\text{E}\left[\tilde{G}_{xx}\right]=P_X$.
When considering individual power spectra, the noise terms and the `signal' terms are indistinguishable, so we therefore do not split the terms up for now.
In this form, we know the classical result that the square of a central normal variable follows a $\chi^2_1$ distribution; also, as the $\tilde{A}$ and $\tilde{B}$ variables are independent, the power spectrum follows a $\chi^2_2$ distribution, scaled by the power in the signal.

\subsubsection{Cross spectrum random variable }
Written out fully, the cross spectrum is a complex random variable $\tilde{G}_{xy} = \tilde{G}_r + \I{} \tilde{G}_i$, where
\begin{equation}
    \begin{split}
        \tilde{G}_r &= H_r (\tilde{A}_s^2 + \tilde{B}_s^2) + \tilde{A}_{u_x} \tilde{A}_{u_y} + \tilde{B}_{u_x} \tilde{B}_{u_y} + \tilde{A}_s \tilde{A}_{u_y} + \tilde{B}_s \tilde{B}_{u_y}  \\
        & \qquad + H_r \tilde{A}_s \tilde{A}_{u_x} + H_i \tilde{A}_s \tilde{B}_{u_x} - H_i \tilde{B}_s \tilde{A}_{u_x} + H_r \tilde{B}_s \tilde{B}_{u_x}\, ,\\
        \tilde{G}_i &= -H_i (\tilde{A}_s^2 + \tilde{B}_s^2) + \tilde{B}_{u_x} \tilde{A}_{u_y} - \tilde{A}_{u_x} \tilde{B}_{u_y} + \tilde{B}_s \tilde{A}_{u_y} - \tilde{A}_s \tilde{B}_{u_y}  \\
        & \qquad + H_r \tilde{A}_s \tilde{B}_{u_x} - H_i \tilde{A}_s \tilde{A}_{u_x} - H_r \tilde{B}_s \tilde{A}_{u_x} - H_i \tilde{B}_s \tilde{B}_{u_x}\, .
    \end{split} 
    \label{eqn:crossspec_real_imag_expanded}
\end{equation}
Both $\tilde{G}_r$ and $\tilde{G}_i$, the cospectrum and the quadrature spectrum, still constitute quadratic forms of random variables. While each still have terms that simplify down to scaled $\chi^2$ distributions, we also have terms that are a product of two uncorrelated normal variables. 
These cross-terms individually produce variance-gamma variables \citepalias[for more information, see][]{Huppenkothen2018}.  Ultimately, many of the constituent terms are innately correlated, so approaching the problem term-by-term is unwise.

\subsection{Distribution parameters}
\label{sec:coherences}
The \citet{Bendat2010} definition of the cross spectrum is 
\begin{equation}
    \langle \tilde{G}_{xy}\rangle = \gamma P_X P_Y
\end{equation}
which is equivalent, but subtly different, to the construction we put forward. We use the brackets $\langle\bullet\rangle$ represent averaging over an infinite signal.
Here, $P_X$ and $P_Y$ are defined by the average power seen in the power spectra, $P_X=\langle \tilde{G}_{xx}\rangle$ and $P_Y=\langle \tilde{G}_{yy}\rangle$; and $\gamma$ is the complex, intrinsic coherence between the two signals.
This formulation is typically more useful for general analysis than the construction we present, and so here we provide a translation.

In this paper, we have constructed our distributions in terms of the underlying signal $P_s$; along with the element of the light curve $x$ that is otherwise uncorrelated with the light curve $y$, $P_{u_x}$; its counterpart $P_{u_y}$; and the transfer function $H=|H|\text{e}^{\I{}\phi_H}$. The more commonly used alternative is to consider the observed light curves, considering directly the power spectra of light curves $x$ and $y$, $P_X$ and $P_y$; the observational noise in each light curve (i.e. Poisson noise) $P_{n_x}$ and $P_{n_y}$ (which can typically be calculated from the count rate and detector characteristics); and the squared intrinsic coherence $\gamma^2$ and the phase lag $\phi$. The formal definition of the squared coherence is \citep{Vaughan1997}
\begin{equation}
    \gamma^2 = \frac{|\langle G_{xy}\rangle|^2}{\langle G_{xx}\rangle \langle G_{yy}\rangle}\,.
    \label{eqn:vaughan_coh}
\end{equation}
When it comes to measuring the coherence using the cross spectrum, there are two ways this is often measured \citep{Vaughan1997,Bendat2010,Uttley2014,Ingram2019}. The first is measuring the raw coherence, that is directly estimating the coherence from the light curves as observed
\begin{equation}
    \hat{g}^2 = \frac{|\hat{G}_{xy}|^2-\hat{b}^2}{\hat{G}_{xx} \hat{G}_{yy}}\,.
    \label{eqn:rawcoherence}
\end{equation}
Here, we indicate quantities that are an estimate from data based upon averaging over finite signal with a hat, i.e. $\hat\bullet$. 
We note the differences between Eqs.~\ref{eqn:vaughan_coh} and \ref{eqn:rawcoherence} that stem from the difference between finite and infinite signal. 
The average over infinite signal allows us to set the observational noise terms $P_{n_x}$ and $P_{n_y}$ to $0$, such that e.g. $\langle \tilde{G}_{xx}\rangle=P_X$; however $\hat{G}_{xx}=P_X+P_{n_x}$.
We also note that $\hat{b}\rightarrow0$ when the averaging tends towards an infinite sample, bringing it in line with Eq.~\ref{eqn:vaughan_coh}.

It is possible to estimate the intrinsic coherence of the signals `at source' $\gamma$, absent of any power caused by measuring the signal.
This is done by subtracting the expected noise from the measured power spectra
\begin{equation}
    \hat{\gamma}^2 = \frac{|\hat{G}_{xy}|^2-\hat{b}^2}{\left(\hat{G}_{xx}-P_{n_x}\right) \left(\hat{G}_{yy}-P_{n_y}\right)}\,.
    \label{eqn:coherence}
\end{equation}
These estimates have two pitfalls. First, the modulus-squared cross spectrum forms a biased estimator, hence the inclusion of the bias term $b^2$ is required (for more details see \citealt{Vaughan1997, Ingram2019}). Second, (as noted by \citet{Uttley2014}) the estimate requires significant averaging such that the uncorrelated noise terms in the estimate of the cross spectrum $\hat{G}_{xy}$ cancel to 0. Therefore, while the underlying concept of the coherence is defined, estimating it from data is more complicated.

\subsubsection{Relation to observables}

We can compare our description of the cross spectrum to the notation used by \citet{Bendat2010} by noticing that there is no mathematical difference between observational noise, and the intrinsically uncorrelated components of the light curves.
This gives the relations

\begin{equation}
    \begin{split}
        H &= \sqrt{\frac{P_Y - P_{n_y}}{P_X - P_{n_x}}} \sqrt{\gamma^2}\text{e}^{-\I{} \phi}=\sqrt{\frac{P_Y - P_{n_y}}{P_X - P_{n_x}}}\gamma, \\
        P_s &= P_X - P_{n_x} , \\
        P_{u_x} &= P_{n_x} , \\
        P_{u_y} &= P_{n_y} + \left(P_Y - P_{n_y}\right) \left(1-\gamma^2\right) .
    \end{split}
    \label{eqn:notation_transform}
\end{equation}
Due to our definition of the transfer function in comparison to the phase lag that will be observed, we have $\phi_H=-\phi$.
It may seem counter-intuitive that there appears to be no element of the uncorrelated signal power in our definition of $P_{u_x}$. However, this is purely a result of our definition of $H$ and $P_s$ which are based upon the power in light curve $x$. This parameterisation is able to produce any amount of coherence between the signals while maintaining the observational noise level and $P_X$.

In Section~\ref{sec:quad_form} we shall define  
\begin{equation}
    \eta\equiv\frac{1}{2}\left(|H|^2P_sP_{u_x}+P_sP_{u_y}+P_{u_x}P_{u_y}\right)\,,
\end{equation}
which will be a useful quantity in the distributions we shall derive.
We will often see $\eta$ appear in the term $|H|^2P_s^2+2\eta$, which, when expanded, is equivalent to $P_XP_Y$. We will also show that $\left|\langle\tilde{G}_{xy}\rangle\right|^2=|H|^2P_s^2$. Therefore, by comparing to Eq.~\ref{eqn:vaughan_coh} we can see that
\begin{equation}
        \frac{|H|^2P_s^2}{|H|^2P_s^2+2\eta}=g^2\,\text{, and }\,
        \frac{2\eta}{|H|^2P_s^2+2\eta}=1-g^2\,.
\end{equation}
These terms will appear in the distributions we shall derive, so we highlight these relations here to provide the physical reason why these are common terms.

\subsection{Vector form of the cross spectrum}
In this paper, we will often write the cross and power spectra as the 4D multi-variate random variable 
\begin{equation}
    \mathbf{\tilde{G}}=
    \begin{pmatrix}
        \tilde{G}_{xx} \\ 
        \tilde{G}_{yy} \\
        \tilde{G}_r \\
        \tilde{G}_i
    \end{pmatrix}\,.
\end{equation} 
When required, we will also consider 2D random variables for just the cross spectrum, the first representation being the real and imaginary components
$\mathbf{\tilde{G}}_{xy}=\left(\tilde{G}_r, \tilde{G}_i\right)'$, but we will also consider the 2D random variable containing the magnitude and phase of the cross spectrum.  The polar form of the cross spectrum can be written as as 
$\tilde{G}_{xy} = \tilde{\rho}_G \text{e}^{\I{}\tilde{\phi}_G}$. We can use $H = |H| \text{e}^{-\I{}\phi}$, to write $\tilde{\Delta}_\phi = \tilde{\phi}_G - \phi$, and we can then make use of the 2D random variable $\mathbf{\tilde{G}}_p=(\tilde{\rho}_G,\tilde{\Delta}_\phi)'$.

We can use similar notation for the random variable that is the average of $N$ realisations of the cross spectrum. We will write 4D random variable of this average with 
\begin{equation}
    \mathbf{\tilde{M}} = \frac{1}{N} \sum^N_n \mathbf{\tilde{G}}_{n}= 
    \begin{pmatrix}
        \tilde{M}_{xx} \\ 
        \tilde{M}_{yy} \\
        \tilde{M}_r \\
        \tilde{M}_i
    \end{pmatrix}  \, , 
\end{equation}
and likewise we can use the 2D marginal form $\mathbf{\tilde{M}}_{xy} = \frac{1}{N} \sum^N_n \mathbf{\tilde{G}}_{xy,n}=(\tilde{M}_r,\tilde{M}_i)'$. Finally, we can also consider the polar form of the $N$-averaged cross spectrum, with the random variable
$\mathbf{\tilde{M}}_p=(\tilde{\rho}_M,\tilde{\Delta}_{M\phi})'$. We highlight that this uses the magnitude and phase of the average of the cross spectrum, 
$\tilde\rho_M = \frac{1}{N} \left| \sum_n^N \mathbf{\tilde{G}}_{xy,n}\right|$ and $\phi_{M\phi} = \arg{\left[ \sum_n^N \mathbf{\tilde{G}}_{xy,n}\right]}$, not the means of the magnitude and phase of the cross spectrum realisations.

\subsection{Simulations}
\label{sec:simulations}

To visualise the distribution of the cross spectrum that we wish to find, we produce three simulations by sampling $10^6$ times from our random variables in a similar manner to \citet{Timmer1995} (the full procedure we use is described in Appendix~\ref{app:simulations}), from which we compute our power spectra and cross spectrum.  These simulations are shown in Fig.~\ref{fig:PxPyGrGi_coh_sims}, where the only difference between the panels is the coherence of the signal between the time series $\mathbf{x}$  and $\mathbf{y}$.  We set the variability powers assuming a normalisation like \citet{Leahy1983} such that the power of the Poisson noise is $P_{n_x}=P_{n_y}=2$, and that the overall power of the time series is $P_X=P_Y=10$; the phase lag between the coherent elements of the signal is $\phi=0.46$ radians. 
We show three values for the intrinsic squared coherence $\gamma^2=0,0.25,1$, and we also provide the equivalent raw squared coherence $g^2$ resulting from our levels of observational noise.

By showing equal-aspect ratio axes, we highlight the intrinsic covariance between the different spectra.
In Fig.~\ref{fig:PxPyGrGi_coh_ave_sims}, we repeat this procedure but each of our $10^6$ samples is itself the average of $N=50$ independent realisations of the same underlying cross spectrum. This demonstrates that, even under averaging, elements of the cross spectrum remain covariant.

\section{Quadratic forms of normal random variables}
\label{sec:quad_forms}

To begin our derivation of the statistical distribution of the cross spectrum, we start by examining the form of both $G_r$ and $G_i$ (Eq.~\ref{eqn:crossspec_real_imag_expanded}). 
We can see that they are both quadratic forms of the 6 dimensional, multivariate central normal random variable, $\mathbf{\tilde{A}}' = \left(\tilde{A}_s, \tilde{B}_s, \tilde{A}_{u_x}, \tilde{B}_{u_x}, \tilde{A}_{u_y}, \tilde{B}_{u_y} \right)$, i.e., we can write both the real and imaginary components of the cross spectrum as $\tilde{G}_r = \mathbf{\tilde{A}}' \mathbf{\Gamma}_r \mathbf{\tilde{A}}$ and $\tilde{G}_i = \mathbf{\tilde{A}}' \mathbf{\Gamma}_i \mathbf{\tilde{A}}$, where $\mathbf{\Gamma}_r$ and $\mathbf{\Gamma}_i$ are real matrices and $\mathbf{\tilde{A}}'$ is the transpose of $\mathbf{\tilde{A}}$.

In general, a quadratic form of normal random variables follows a generalised $\chi^2$ distribution which does not have an analytic PDF \citep{IMHOF1961}; the two dimensional form for two such joint distributions is therefore not expected to exist in the general form.  However, we can start by considering some general properties. 
For a multivariate normal vector $\mathbf{\tilde{Z}}\sim \mathcal{N}\left(\mathbf{m}_z, \mathbf{\Sigma}_z\right)$ mixed by a symmetric matrix $\mathbf{\Gamma}$ \citep{Rencher2008}
\begin{equation}
    \begin{split}
        \text{E}\left[\mathbf{\tilde{Z}}'\mathbf{\Gamma}\mathbf{\tilde{Z}}\right] &= \text{tr}\left[ \mathbf{\Sigma}_z\mathbf{\Gamma}\right] + \mathbf{m}_z'\mathbf{\Gamma}\mathbf{m}_z, \\
        \text{Var}\left[\mathbf{\tilde{Z}}'\mathbf{\Gamma}\mathbf{\tilde{Z}}\right] &= 2\text{tr}\left[ \mathbf{\Sigma}_z\mathbf{\Gamma}\mathbf{\Sigma}_z\mathbf{\Gamma}\right] + 4\mathbf{m}_z'\mathbf{\Gamma}\mathbf{\Sigma}_z\mathbf{\Gamma}\mathbf{m}_z,
    \end{split}
    \label{eqn:quad_form_ex_var}
\end{equation}
and for two different quadratic forms of the same random normal vector $\mathbf{\tilde{Z}}$ \citep{Graybill1983} 
\begin{equation}
    \text{Cov}\left[\mathbf{\tilde{Z}}'\mathbf{\Gamma_1}\mathbf{\tilde{Z}}, \mathbf{\tilde{Z}}'\mathbf{\Gamma}_2\mathbf{\tilde{Z}}\right] = 2\text{tr}\left[ \mathbf{\Sigma}_z\mathbf{\Gamma}_1\mathbf{\Sigma}_z\mathbf{\Gamma}_2\right] + 4\mathbf{m}_z'\mathbf{\Gamma}_1\mathbf{\Sigma}_z\mathbf{\Gamma}_2\mathbf{m}_z.
    \label{eqn:quad_form_cov}
\end{equation}

For the rest of this section, we will use this quadratic form to re-derive the results shown in Section 9.1 of \citet{Bendat2010}.

\subsection{The quadratic form of the cross spectrum}
\label{sec:quad_form}
Considering the above notation, the mean vector $\mathbf{m_A}$ and covariance matrix $\mathbf{\Sigma_A}$ of our random variable $\mathbf{\tilde{A}}$ come directly from Eq.~\ref{eqn:normal_variables}, noting that the six variables are completely uncorrelated.
Looking at Eq.~\ref{eqn:crossspec_real_imag_expanded}, we can simply write down the mixing matrices $\mathbf{\Gamma}_r$ and $\mathbf{\Gamma}_i$ (Eqs.~\ref{eqn:mix_mat_Gr} and \ref{eqn:mix_mat_Gi} in Appendix~\ref{app:mix_mat}).
With these, we can use Eq.~\ref{eqn:quad_form_ex_var} to find the expectations of the random variables $\tilde{G}_r$ and $\tilde{G}_i$ in Eq.~\ref{eqn:crossspec_real_imag_expanded}
\begin{equation}
    \begin{split}
    \text{E}\left[\tilde{G}_r\right] &= H_r P_s \, \textrm{ and} \\
    \text{E}\left[\tilde{G}_i\right] &= -H_i P_s\,,
    \end{split}
    \label{eqn:mean_GrGi}
\end{equation}
and the variances
\begin{equation}
    \begin{split}
         \text{Var}\left[\tilde{G}_r\right] &= \left(H_r P_s\right)^2 + \eta \, \textrm{ and} \\
         \text{Var}\left[\tilde{G}_i\right] &= \left(H_i P_s\right)^2 + \eta \,,
    \end{split}
    \label{eqn:var_GrGi}
\end{equation}
where we define $\eta\equiv\frac{1}{2}\left(|H|^2P_sP_{u_x}+P_sP_{u_y}+P_{u_x}P_{u_y}\right)$ to be concise.

Further, we can use Eq.~\ref{eqn:quad_form_cov} to find the intrinsic covariance between $\tilde{G}_r$ and $\tilde{G}_i$,
\begin{equation}
         \text{Cov}\left[\tilde{G}_r, \tilde{G}_i\right] = - H_r H_i P_s^2\,,
\end{equation}
agreeing with results of \citet{Bendat2010} (see their table 9.2).

\subsection{The quadratic form of the power spectra}

We can also write our two power spectra random variables $\tilde{G}_{xx}$ and $\tilde{G}_{yy}$ in a similar notation, however this time expanding upon the random Fourier amplitudes $\tilde{A}_{\{x,y\}}$ and $\tilde{B}_{\{x,y\}}$
\begin{equation}
    \begin{split}
        \tilde{G}_{xx} &= \tilde{A}_s^2 + \tilde{B}_s^2 + \tilde{A}_{u_x}^2 + \tilde{B}_{u_x}^2 + 2 \tilde{A}_s \tilde{A}_{u_x} + 2 \tilde{B}_s \tilde{B}_{u_x}\, ,\\
        \tilde{G}_{yy} &= |H|^2 \left(\tilde{A}_s^2 + \tilde{B}_s^2\right) + \tilde{A}_{u_y}^2 + \tilde{B}_{u_y}^2 + 2 H_r \tilde{A}_s \tilde{A}_{u_y} + 2 H_i \tilde{A}_s \tilde{B}_{u_y} \\
        & \qquad - 2 H_i \tilde{B}_s \tilde{A}_{u_y} + 2 H_r \tilde{B}_s \tilde{B}_{u_y}\, .
    \end{split} 
    \label{eqn:powerspec_expanded}
\end{equation}
It is obvious that writing this explicitly cannot change the known result that the power spectra follow a $\chi^2_2$ distribution. Pushing forward to find the mixing matrices $\mathbf{\Gamma}_X$ and $\mathbf{\Gamma}_Y$ (Eqs.~\ref{eqn:mix_mat_PX} and \ref{eqn:mix_mat_PY} in Appendix~\ref{app:mix_mat}),
we apply Eq.~\ref{eqn:quad_form_cov} to show that the mean and variance of the power spectra random variables are as expected
\begin{equation}
    \begin{split}
    \text{E}\left[\tilde{G}_{xx}\right] &= P_s + P_{u_x} \,, \\
    \text{E}\left[\tilde{G}_{yy}\right] &= |H|^2 P_s + P_{u_y}\, \textrm{ and} \\ 
    \text{Var}\left[\tilde{G}_{xx}\right] &= \left(P_s + P_{u_x}\right)^2  \,, \\
    \text{Var}\left[\tilde{G}_{yy}\right] &= \left(|H|^2 P_s + P_{u_x}\right)^2\,.
    \end{split}
\end{equation}

We can also compute the covariance between the two power spectra, and with the cross spectrum, using Eq.~\ref{eqn:quad_form_cov},
\begin{equation}
    \begin{split}
        \text{Cov}\left[\tilde{G}_{xx}, \tilde{G}_{yy}\right] &= |H|^2 P_s^2 \,, \\
        \text{Cov}\left[\tilde{G}_{xx}, \tilde{G}_r\right] &= H_r P_s (P_s + P_{u_x}) \,, \\
        \text{Cov}\left[\tilde{G}_{xx}, \tilde{G}_i\right] &= -H_i P_s (P_s + P_{u_x}) \,, \\
        \text{Cov}\left[\tilde{G}_{yy}, \tilde{G}_r\right] &= H_r P_s (|H|^2 P_s + P_{u_y}) \,, \\
        \text{Cov}\left[\tilde{G}_{yy}, \tilde{G}_i\right] &= -H_i P_s (|H|^2 P_s + P_{u_y}) \,.
    \end{split}
\end{equation}

\subsection{Joint consideration of power spectra and cross spectrum}

If we consider the power spectra and the cross spectra as the multivariate random variable $\mathbf{\tilde{G}}$, we can write the mean vector and the covariance matrix as
\begin{equation}
    \begin{split}
        \mathbf{m_{G}} &= 
          \begin{pmatrix}
            P_s + P_{u_x} \\ |H|^2 P_s + P_{u_y} \\ H_r P_s \\ -H_i P_s
          \end{pmatrix} \, \text{, and }\\
        \mathbf{\Sigma}_\mathbf{G} &= 
          \begin{pmatrix}
               0 & -2\eta &    0 &    0 \\ 
          -2\eta &      0 &    0 &    0 \\
               0 &      0 & \eta &    0 \\ 
               0 &      0 &    0 & \eta \\
          \end{pmatrix} + \mathbf{m_{G}}' \mathbf{m_{G}} \,.
    \end{split}
    \label{eqn:av_mean_cov}
\end{equation}

\subsection{Fitting an averaged cross spectrum}

While the full probability distribution function of a single realisation of the cross spectrum is complicated, we can start by considering the case where we average many realisations together. 
Following \citet{vanderKlis1989}, we can consider two ways of selecting bins to average: a number of consecutive frequency bins, or bins selected from the same frequency from the cross spectrum computed from multiple segments of a light curve. By using both of these two options, it can be tuned to optimise for the range of frequencies and the frequency resolution needed. The crucial assumption here is that all the bins chosen to be averaged must be independent measurements of the same cross spectrum. For the first case, this means that consecutive frequencies have the same properties (which may not be true near narrow features); while the second case requires the cross spectrum to be stationary, i.e. the light curves are driven by the same underlying variability process across all segments.

Provided many realisations are averaged, the central limit theorem becomes a valid approximation. If we designate $\mathbf{\tilde{M}}$ as the average of $N$ independent realisations of $\mathbf{\tilde{G}}$, the central limit theorem indicates that $\mathbf{\tilde{M}}$ approximately follows a normal distribution with the mean vector $\mathbf{m_{G}}$ and covariance matrix $\frac{1}{N}\mathbf{\Sigma}_\mathbf{G}$.

\subsubsection{Sufficient averaging}

When the spectra are sufficiently averaged that their joint distribution is adequately approximated as a multivariate normal distribution, we can use maximum likelihood estimation to estimate the best fitting parameters.

When $\mathbf{\tilde{M}}$ is the average of a sample of $N$ independent realisations of the cross spectrum, we can fit a set of parameters $\theta=\left\{P_s,P_{u_x},P_{u_y},H\right\}$ which predict a distribution mean $\mathbf{m}_\theta$ and covariance $\mathbf{\Sigma}_\theta$\footnote{$\mathbf{\Sigma}_\theta$ here is the covariance of the underlying, unaveraged distribution.}. The likelihood of the parameter set is given by
\begin{equation}
    \mathcal{L}(\theta) = \frac{N^{k/2}}{(2\pi)^{k/2} \left(\det{\mathbf{\Sigma}_\theta}\right)^{1/2}}\exp{\left(-\frac{N}{2} (\mathbf{\tilde{M}} - \mathbf{m}_\theta)'\mathbf{\Sigma}_\theta^{-1}(\mathbf{\tilde{M}} - \mathbf{m}_\theta)\right)}\,,
    \label{eqn:averaged_likelihood}
\end{equation}
where $k$ is the number of dimensions being fit, and $\mathbf{\Sigma}_\theta^{-1}$ is the inverse matrix of $\mathbf{\Sigma}_\theta$. In the case where both power spectra, the cospectrum, and the quadrature spectrum are all fit, then $k=4$; whereas, if only the cospectrum and the quadrature spectrum are jointly fit (i.e. when only the cross spectrum itself is considered) then $k=2$.

Caution must be taken when using this approach, as innately there is an attempt to fit five distribution parameters to four observables\footnote{Here, fitting the five parameters $P_s$, $P_{u_x}$, $P_{u_y}$, $H_r$, and $H_i$ to the two power spectra, the cospectrum, and the quadrature spectrum.} 
However, this can be mitigated when assumptions can be made about parameters(for example, knowledge of the expected Poisson noise), or using models can be fit to multiple frequencies simultaneously.

\subsubsection{Fit statistic}

A fit statistic to minimise can be simply found by looking at the log-likelihood $\ln\mathcal{L}$. 
We produce a relevant statistic for fitting the cross-spectrum by taking $-2\ln \mathcal{L}$
\begin{equation}
    -2\ln\mathcal{L} = N(\mathbf{\tilde{M}} - \mathbf{m}_\theta)'\mathbf{\Sigma}_\theta^{-1}(\mathbf{\tilde{M}} - \mathbf{m}_\theta) + \det{\mathbf{\Sigma}_\theta}\,,
    \label{eqn:loglikelihoodStatistic}
\end{equation}
which would reduce to typical $\chi^2$ statistics if $\mathbf{\Sigma}_\theta$ were the identity matrix, i.e. if there were no covariance between any of the power, co, or quadrature spectra.  While this statistic does not follow a $\chi^2$ distribution, it is additive with the $\chi^2$ statistic that could come from, e.g. an accompanying flux-energy spectral fit.

This is similarly true for any statistic derived from $-2\ln\mathcal{L}$, such as Cash statistics \citep{Cash1979}.
While the combination of these statistics is valid for maximum likelihood estimation methods, we remind the reader that the combination of Eq.~\ref{eqn:loglikelihoodStatistic} with $\chi^2$ (or any other statistic) means that typical statistical tests, such as $\Delta\chi^2$ methods for estimating parameter uncertainties will not be valid. The statistic Eq.~\ref{eqn:loglikelihoodStatistic} will itself be distributed by the generalised chi-squared distribution, which has no closed form probability distribution function (PDF) or cumulative distribution function (CDF).

Within the field of X-ray spectral-timing, it is not unusual for an uneven binning scheme to be used, e.g. logarithmically binned in frequency space compared to the native linear frequency bins of the DFT.  In such a case, we denote the index of the re-binned frequencies as $j'$, and the total fit statistic becomes
\begin{equation}
    \sum_{j'}     N_{j'}(\mathbf{\tilde{M}}_{j'} - {\mathbf{m}_\theta}_{j'})'{\Sigma_\theta}_{j'}^{-1}(\mathbf{\tilde{M}}_{j'} - {\mathbf{m}_\theta}_{j'}) + \det{{\Sigma_\theta}_{j'}}\,,
\end{equation}
where each frequency bin in the re-binned cross spectrum is the average of $N_{j'}$ realisations; as before $\mathbf{\tilde{M}}_{j'}$ is the measured mean of the cross spectrum of the bin $j'$, with ${\mathbf{m}_\theta}_{j'}$ and ${\Sigma_\theta}_{j'}$ also being the predicted mean and covariance of the cross spectral components in the un-averaged case.

\section{Distribution of a single cross spectrum}
\label{sec:crossspectrum_dist}

The preceding sections discussed the typical use-case of an averaged cross spectrum.  However, when it is impractical to produce an averaged cross spectrum, it is useful to have the distribution for a single case.

We approach this in a form related to the moment generating function used by \citetalias{Huppenkothen2018}, but instead use the characteristic function.

\subsection{The characteristic function}

The characteristic function (CF) is related to the Fourier transform of the PDF.  For a general D-dimensional random vector $\mathbf{\tilde{Z}}$ with the PDF $f_\mathbf{Z}(\mathbf{Z})$, the CF is defined as
\begin{equation}
    \varphi_\mathbf{Z}(\mathbf{t}) =\text{E}\left[\text{e}^{\I\mathbf{t}'\mathbf{Z}}\right]
     \equiv \oint f_\mathbf{Z}(\mathbf{Z}) \text{e}^{\I\mathbf{t}'\mathbf{Z}} \dd{}\mathbf{Z} \,.
\end{equation}

When considering the vector $\mathbf{G}$, we can write the dot product as 
\begin{equation}
    \mathbf{t}'\mathbf{G} = \sum^D_d t_d G_d = \mathbf{A}'\left(\sum^D_d t_d \mathbf{\Gamma}_d \right) \mathbf{A}\,.
\end{equation}
Therefore, we can express this itself as a quadratic form of normal variables.  Following \citet{Singull2012, Slichenko2014}, we can write the CF of $\mathbf{G}$ as
\begin{equation}
    \begin{split}
        \varphi_\mathbf{G}(\mathbf{t}) &= \oint f_\mathbf{A}(\mathbf{A}) \exp{\left[\I\mathbf{A}'\left(\sum_d t_d \mathbf{\Gamma}_d \right) \mathbf{A}\right]} \dd{}\mathbf{A} \\
        & = \frac{1}{(2\pi)^3\sigma_s^2\sigma_{u_x}^2\sigma_{u_y}^2}\oint \exp{\left[- \frac{1}{2}\mathbf{A}'\left(\Sigma_A^{-1} - 2\I\sum_d t_d\mathbf{\Gamma}_d \right) \mathbf{A}\right]} \dd{}\mathbf{A} \,,
    \end{split}
\end{equation}
where we have inserted the probability distribution of the multivariate normal variable $\mathbf{A}$.  Using the normalisation of a multivariate normal distribution, the CF is
\begin{equation}
    \begin{split}
    \varphi_\mathbf{G}(\mathbf{t}) &= \frac{1}{\sigma_s^2\sigma_{u_x}^2\sigma_{u_y}^2}\text{det}\left[\mathbf{\Sigma}_A^{-1} - 2\I\sum_n t_n \mathbf{\Gamma}_n \right]^{-\frac12} \\
    &=\frac{1}{1 - \I{} \mathbf{m}'\mathbf{t} + \frac{1}{2}\mathbf{t}'\mathbf{\Lambda}\mathbf{t}}\,,
    \end{split}
    \label{eqn:G_char_func}
\end{equation}
where
\begin{equation}
      \mathbf{m} = 
      \begin{pmatrix}
        P_s + P_{u_x} \\ |H|^2 P_s + P_{u_y} \\ H_r P_s \\ -H_i P_s
      \end{pmatrix} \, \text{, and }
      \mathbf{\Lambda} = 
      \begin{pmatrix}
           0 & -2\eta &    0 &    0 \\ 
      -2\eta &      0 &    0 &    0 \\
           0 &      0 & \eta &    0 \\ 
           0 &      0 &    0 & \eta \\
      \end{pmatrix} \,,
\end{equation}
which are directly related to the mean and covariance of the distribution (c.f. Eq.~\ref{eqn:av_mean_cov}).

\subsection{Asymmetric Laplace Distribution}

To understand the probability distribution, we can compare the CF of $\mathbf{G}$ to the CF of a multivariate asymmetric Laplace (AL) distribution \citep{Kotz2001}
\begin{equation}
    \varphi_{\text{AL}}(\mathbf{t}) =
    \frac{1}{1 - \I{} \mathbf{m}_{\text{AL}}'\mathbf{t} + \frac{1}{2}\mathbf{t}'\mathbf{\Lambda}_{\text{AL}}\mathbf{t}}\,,
\end{equation}
which is identical to the form shown in Eq.~\ref{eqn:G_char_func}, with one key exception: $\mathbf{\Lambda}_{\text{AL}}$ is required to be a nonnegative definite matrix.  Unfortunately, our $\mathbf{\Lambda}$ does not meet that requirement; however when reducing the problem to just the cross spectrum (i.e., dropping the two power spectra), we recover a matrix that meets the requirement.  Therefore, the cospectrum and the quadrature spectrum together are distributed as a bi-variate Asymmetric Laplace distribution.

\subsubsection{Probability Density Function}

\citet{Kotz2001} gives the general form of the PDF for the bi-variate\footnote{Eq.~6.5.3 of \citet{Kotz2001}, where for two dimensions we have simplified their general form using $d=2$ and therefore $v=0$.} Asymmetric Laplace variable $\mathbf{\tilde{Z}}$
\begin{equation}
    f_{\text{AL}}(\mathbf{Z}) = \frac{\text{e}^{\mathbf{Z}'\mathbf{\Lambda}^{-1}_\text{AL}\mathbf{m_{\text{AL}}}}}{\pi \sqrt{\det\mathbf{\Lambda}_{\text{AL}}}}K_0\left(\sqrt{(2+\mathbf{m}_{\text{AL}}'\mathbf{\Lambda}_{\text{AL}}^{-1}\mathbf{m}_{\text{AL}}) (\mathbf{Z}'\mathbf{\Lambda}_{\text{AL}}^{-1}\mathbf{Z})}\right) ,
    \label{eqn:vector_general_pdf}
\end{equation}
where $K_0$ is the zeroth-order modified Bessel function of the second kind.

For our case, where we are using the 2D random variable $\mathbf{\tilde{G}}_{xy}$ with
\begin{equation}
    \mathbf{m}_{xy}=
       \begin{pmatrix}
            H_rP_s \\
           -H_iP_s
       \end{pmatrix}\,\text{ and }\,
    \mathbf{\Lambda}_{xy}=
       \begin{pmatrix}
           \eta &    0 \\
              0 & \eta   
       \end{pmatrix}\,,
\end{equation}
the PDF is
\begin{equation}
    f(G_r, G_i) = \frac{1}{\pi \eta}\text{e}^{\eta^{-1} (H_r G_r - H_i G_i) P_s} K_0\left(\frac{\sqrt{|H|^2 P_s^2 + 2\eta}}{\eta} |G|\right) ,
    \label{eqn:PDF}
\end{equation}
where $|G|\equiv\sqrt{G_r^2+G_i^2}$.
It therefore follows that the random variable $\tilde{\mathbf{G}}_{xy}$ which follows this distribution has the expectation and covariance matrix
\begin{equation}
    \begin{split}
      \text{E}\left[\mathbf{\tilde{G}_{xy}}\right] &=
      \begin{pmatrix}
          H_r P_s \\ 
          -H_i P_s
      \end{pmatrix} \, \text{, and }\\
      \text{Cov}\left[\mathbf{\tilde{G}_{xy}}\right] &=  
      \begin{pmatrix} 
         \eta + H_r^2P_s^2  &      - H_rH_iP_s^2 \\  
              - H_rH_iP_s^2 & \eta + H_i^2P_s^2  \\
      \end{pmatrix} \,.
    \end{split}
\end{equation}

\subsection{Distributions of the cospectrum and quadrature spectrum}
\label{sec:co_quad}

The cospectrum and quadrature spectrum are the real and imaginary parts of the cross spectrum.
We can find the CF of these either by marginalising the distribution of the cross spectrum directly from Eq.~\ref{eqn:G_char_func} (see Appendix~\ref{app:marginalising_char}), or computing them directly in an equivalent way.  We find the CFs of the cospectrum and quadrature spectrum
\begin{equation}
         \varphi_r(t_r) = \frac{1}{1+\frac{\eta}{2}t_r^2 - \I{}P_s H_r t_r} \, \text{ and } \,
         \varphi_i(t_i) = \frac{1}{1+\frac{\eta}{2}t_i^2 + \I{}P_s H_i t_i}\,,
         \label{eqn:real_imag_CFs}
\end{equation}
which, through the same comparison, are the CFs of single-variable asymmetric Laplace distributions located at $0$~\citep{Kotz2001}. Therefore, the PDFs of the random variables $\tilde{G_r}$ and $\tilde{G_i}$ are
\begin{equation}
    \begin{split}
        f_r(G_r) &= \frac{ \text{e}^{\left[ \frac{1}{\eta} \left( H_r P_s G_r - |G_r| \sqrt{H_r^2P_s^2 + 2\eta } \right)  \right]}}{ \sqrt{H_r^2P_s^2 + 2\eta}} \, \text{ and } \\
        f_i(G_i) &= \frac{ \text{e}^{\left[ \frac{1}{\eta} \left(-H_i P_s G_i - |G_i| \sqrt{H_i^2P_s^2 + 2\eta } \right)  \right]}}{ \sqrt{H_i^2P_s^2 + 2\eta}} \, .
     \end{split}
     \label{eqn:real_imag_PDFs}
\end{equation}
As a simple check, we can see that these are consistent with known distributions.
In the case where there is no uncorrelated signal (i.e. $\eta\rightarrow0$), the limit of these functions is a $\chi^2_{k=2}$ distribution as one would expect for a power spectrum; likewise the limit with no correlated signal (i.e. $P_s\rightarrow0$), the limit is a Laplace distribution as calculated by \citetalias{Huppenkothen2018}.

It is important to remember that $\tilde{G}_r$ and $\tilde{G}_i$ are not independent even when $H=0$, and we can clearly see $f(G_r, G_i)\neq f_r(G_r)f_i(G_i)$ (see \citetalias{Huppenkothen2018} for a discussion on the cospectrum in this case).
The mean and variance of these distributions trivially follow from Eqs.~\ref{eqn:mean_GrGi} \& \ref{eqn:var_GrGi}.

\subsection{Magnitude and phase of the cross spectrum}

While working with the co and quadrature spectra is useful in many cases, typically the magnitude and phase of the cross spectrum have more physical meaning.  Here we present some distributions for the statistics of these quantities.

We use a coordinate transform of the Cartestian representation of the PDF in Eq.~\ref{eqn:PDF} such that it uses the polar form of $\tilde{G}_{xy} = \tilde{\rho}_G \text{e}^{\I{}\tilde{\phi}_G}$.
We then find the PDF for the 2D random variable $\mathbf{\tilde{G}}_p$ 
\begin{equation}
    f_p\left(\rho_G ,\Delta_\phi\right) = \frac{\rho_G}{\pi \eta}\text{e}^{\frac{|H|P_s}{\eta} \rho_G  \cos{\Delta_\phi}} K_0\left( \frac{\sqrt{|H|^2 P_s^2+2\eta}}{\eta} \rho_G \right) .
    \label{eqn:polar_PDF}
\end{equation}

\subsubsection{Marginalised amplitude of the cross spectrum}
Using the polar form of the cross spectrum, we can consider how the magnitude of the cross spectrum $\tilde{\rho}_G$ is distributed. We see that Eq.~\ref{eqn:polar_PDF} has a simple dependence on $\Delta_\phi$ (and therefore $\phi_G$), so the marginalisation is straight-forward integration.  Calculating, we find
\begin{equation}
    f_{\rho_G}(\rho_G) = \frac{2 \rho_G}{\eta} I_0\left( \frac{|H| P_s}{\eta} \rho_G \right) K_0\left( \frac{\sqrt{|H|^2 P_s^2+2\eta}}{\eta} \rho_G \right) ,
    \label{eqn:mag_PDF}
\end{equation}
where $I_0$ and $K_0$ are the modified Bessel functions of the first and second kind respectively, each of order 0.

The expectation and variance of the magnitude of the cross spectrum are
\begin{equation}
    \begin{split}
        \text{E}\left[\tilde{\rho}_G\right] &= \frac{(|H|^2 P_s^2 + 2\eta) \mathcal{E}\left[ \frac{|H|^2 P_s^2}{|H|^2 P_s^2 + 2\eta}\right] - \eta\mathcal{K}\left[ \frac{|H|^2 P_s^2}{|H|^2 P_s^2 + 2\eta}\right]}{\sqrt{|H|^2 P_s^2 + 2\eta}} \\
        \text{Var}\left[\tilde{\rho}_G\right] &= 2(|H|^2 P_s^2 + \eta) - \left(\text{E}\left[\tilde{\rho}_G\right]\right)^2 \,
    \end{split}
    \label{eqn:execpted_magnitude}
\end{equation}
where $\mathcal{K}[\ldots]$ and $\mathcal{E}[\ldots]$ are respectively the complete elliptic integrals of the first and second kind.

\subsubsection{A note about biased estimators}
\label{sec:biased_magnitude}
Above, we have given the expectation and variance of the distribution of the amplitude of the cross spectrum. In this case, we see that
\begin{equation}
    \text{E}\left[\tilde{\rho}_G\right]\neq |H|P_s
\end{equation}
as one might have na\"ively expected. Therefore, estimating the magnitude of the cross spectrum directly from the data leads to a bias (as appears in Eqs.~\ref{eqn:rawcoherence} and \ref{eqn:coherence}).  This can also be seen in the discussion within Appendix~\ref{app:BP10errs}.  While the bias term can be estimated, we would instead recommend using maximum likelihood estimation with Eq.~\ref{eqn:mag_PDF}. 

\subsubsection{Marginalised phase of the cross spectrum}
Similarly, marginalising Eq.~\ref{eqn:polar_PDF} over $\rho_G$ gives the PDF
\begin{equation}
\begin{split}
f_{\Delta_\phi}(\Delta_\phi) &=  \frac{\eta}{\pi \left(|H|^2 P_s^2 \sin^2\Delta _{\phi} +2 \eta\right)^\frac{3}{2}}\Bigg( \sqrt{|H|^2 P_s^2\sin^2\Delta _{\phi}+2 \eta} \; + \\
 & \qquad\qquad\qquad  |H| P_s \cos\Delta _{\phi}  \arccos\left[\frac{-|H| P_s \cos\Delta
_{\phi}}{\sqrt{|H|^2 P_s^2+2 \eta}}\right]\Bigg)
\,.
\end{split}
\label{eqn:ang_PDF}
\end{equation}
It is important to remember that this is the PDF of the random variable $\tilde{\Delta}_\phi = \tilde{\phi}_G - \phi$, however it can be trivially re-centered to get
$f_{\phi_G}(\phi_G)$.

When considering the angular period $-\pi\leq\Delta_\phi<\pi$, the expectation of $\tilde{\Delta}_\phi = \tilde{\phi}_G - \phi$ is trivially 0 as the PDF is symmetric in its domain $\Delta_\phi = [-\pi,\pi)$.
However, the distribution is periodic, and so it does not necessarily make sense to use the expectation and variance for $\tilde{\Delta}_G$.  We recommend using the full PDF of the cross spectrum.

\subsection{Comparison to simulations}

\begin{figure}
    \centering
    \includegraphics[]{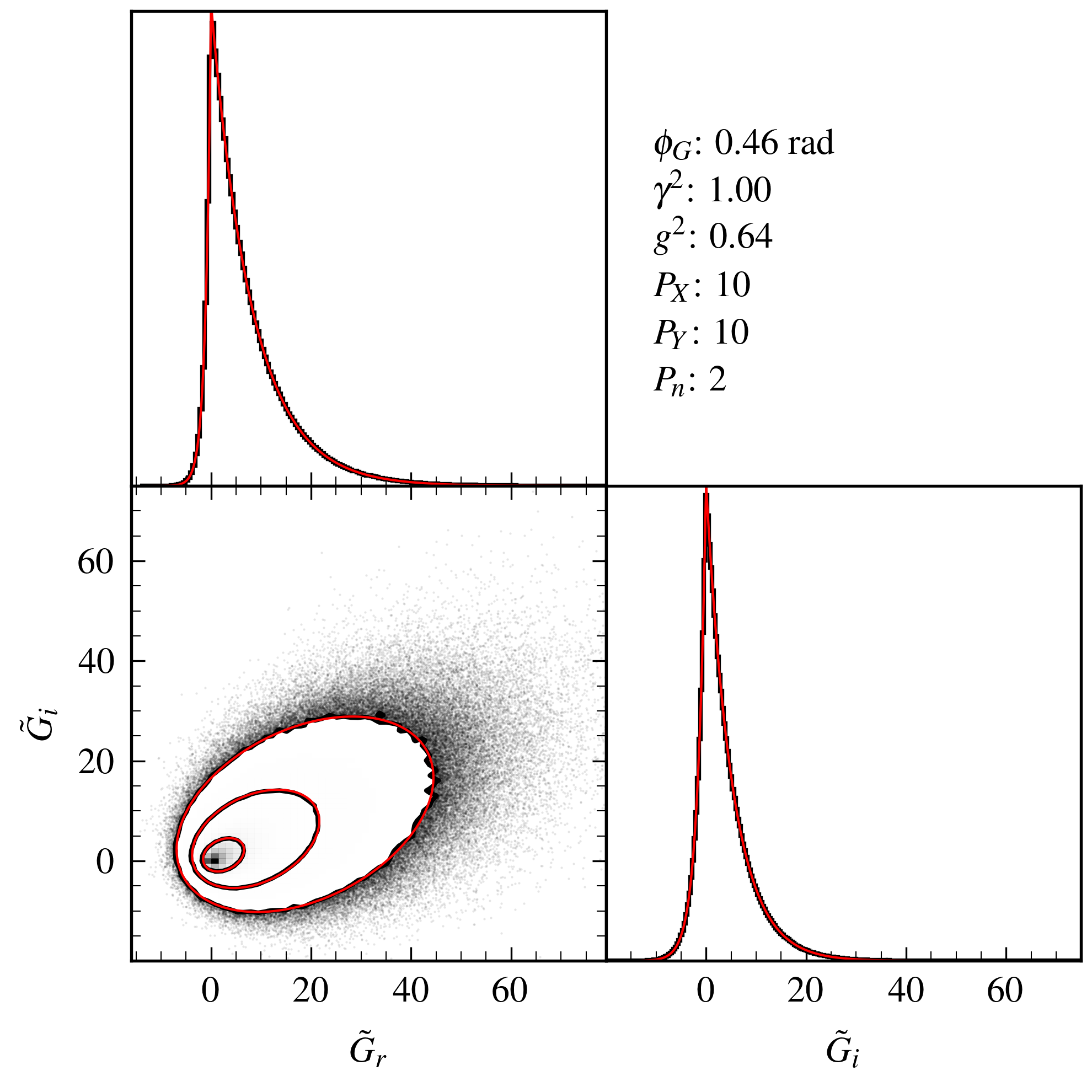}
    \caption{A simulation of 4,000,000 realisations of just the co and quadrature spectra (as described in Appendix~\ref{app:simulations}). The 2D histogram has the $1,2,3\,\sigma$ contours from both the simulation (in black) and calculated from the PDF Eq.~\ref{eqn:PDF} (in red). The marginalised histograms also are also compared to their expected PDFs in Eq.~\ref{eqn:real_imag_PDFs}.}
    \label{fig:Gr_Gi_comparison}
\end{figure}

\begin{figure}
    \centering
    \includegraphics[]{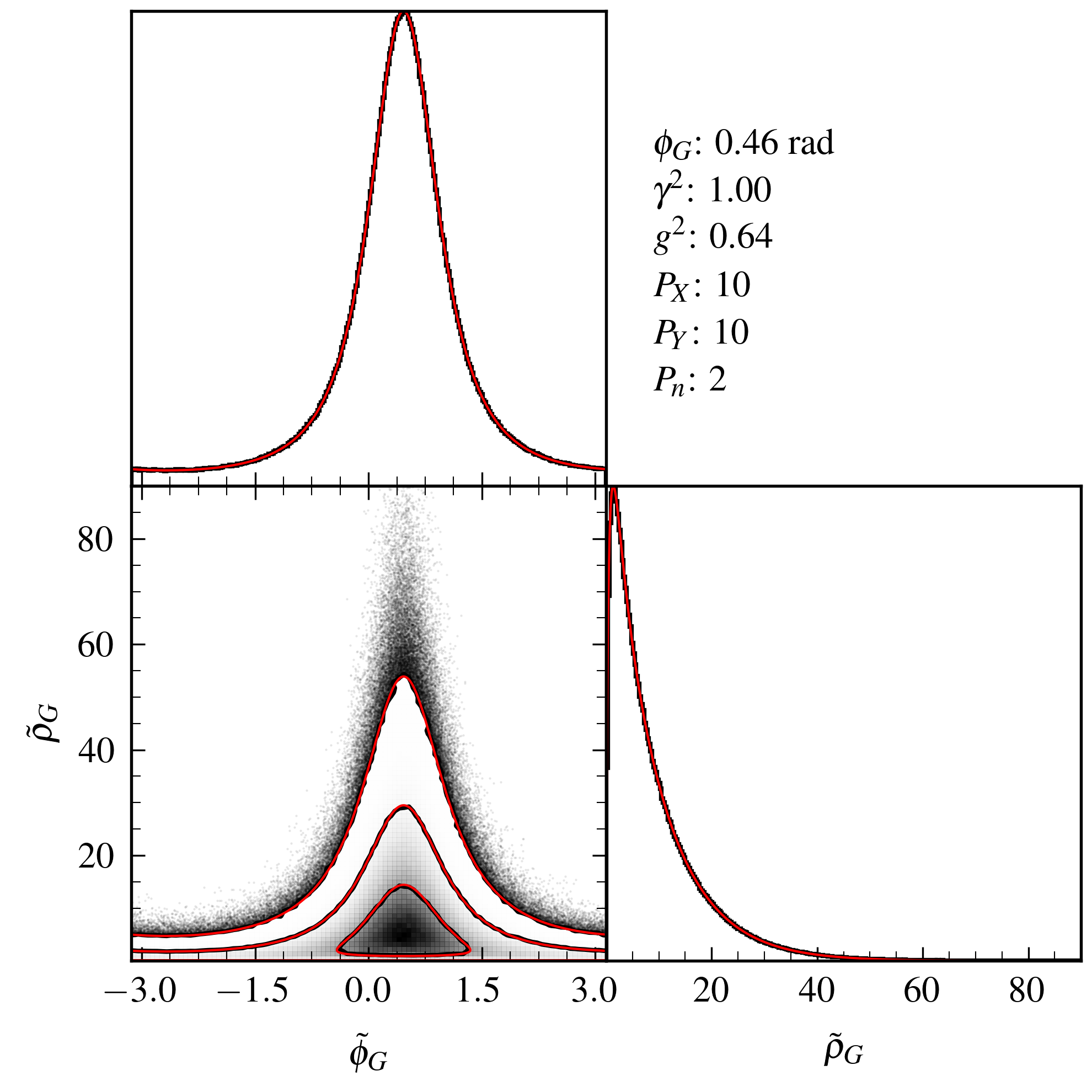}
    \caption{The same simulation as in Fig.~\ref{fig:Gr_Gi_comparison} with 4,000,000 realisations, showing only the joint distribution of the magnitude and phase of the cross spectrum. The 2D histogram has the same contours from the simulation (in black) and from the PDF Eq.~\ref{eqn:polar_PDF} (in red). The marginalised histograms also are also compared to their expected PDFs in Eqs.~\ref{eqn:mag_PDF} and \ref{eqn:ang_PDF}.}
    \label{fig:mag_arg_comparison}
\end{figure}

To experimentally verify of our results, we produce similar simulation as described in Section~\ref{sec:simulations}.  
As before, these simulations are set up such that the time series $\mathbf{x}$ and $\mathbf{y}$ would have a power spectra with RMS $10$, including a noise contribution of $2$, which aligns with a signal that contains Poisson noise when calculated in a \citet{Leahy1983} normalisation.  The phase lag is set as $0.46$~rad, while the intrinsic signal between the signal in the two sources is $1$.  This corresponds to $H=\frac{1}{\sqrt{1.25}}(1-0.5\I{})$, $P_s=8$, and $\eta=18$ in our notation.
In Fig.~\ref{fig:Gr_Gi_comparison} we show the simulation compared to both Eq.~\ref{eqn:PDF} and Eq.~\ref{eqn:real_imag_PDFs}; we also show an identical simulation in Fig.~\ref{fig:mag_arg_comparison} which is compared to Eqs.~\ref{eqn:polar_PDF}, \ref{eqn:mag_PDF}, and \ref{eqn:ang_PDF}.

\section{Low-N averaging}
\label{sec:lowNav}

It is common in the astrophysics community to average a number of cross spectral bins together, either computed from different segments of light curves, or from neighbouring frequencies. Here, we consider the case where the number of averaged realisations of the cross spectrum may not be sufficient to assume Gaussian statistics. 
For the power spectra, the N-averaged result is well known: they follow a $\chi^2$ with $2N$ degrees of freedom. For the cross spectrum, the mean and covariances of averaged quantities trivially follow from that of their underlying distributions.

To consider our random variable $\mathbf{\tilde{M}}_{xy}=\frac{1}{N}\sum_n^N \mathbf{\tilde{G}}_{xy,n}$, we will consider the sum of the scaled random variables $\left\{\frac{1}{N}\mathbf{\tilde{G}}_{xy}\right\}_n$. As this is just a linear scaling, these random variables are the same Asymmetric multivariate Laplace random variables as before, with distribution parameters $\mathbf{m}_{\frac{1}{N}xy}=N^{-1}\mathbf{m}_{xy}$ and $\mathbf{\Lambda}_{\frac{1}{N}xy}=N^{-2}\mathbf{\Lambda}_{xy}$. On the assumption that we are summing completely independent random variables, we can write the CF of $\mathbf{\tilde{M}}_{xy}$ as $\varphi_{M,N}=\left(\varphi_{\frac{1}{N}G}\right)^N$.
The resulting CF is that of a generalised asymmetric Laplace distribution \citep{Kotz2001, Kozubowski2013}. We write the N-averaged cross spectrum as the vector $\mathbf{\tilde{M}}_{xy}$, and this random variable follows the PDF
\begin{equation}
    \begin{split}
        f_{\mathbf{M},N}(M_r, M_i) = \frac{N^{N+1}|M|^{N-1}}{\pi \eta \Gamma(N) } &\left({|H|^2 P_s^2 +2\eta} \right)^\frac{1-N}{2} 
        \text{e}^{\frac{N P_s}{\eta}(H_r M_r - H_i M_i)}  \\
        &\times K_{N-1}\left(\frac{N}{\eta}\sqrt{|H|^2 P_s^2+2\eta }|M|\right) ,
    \label{eqn:N_av_PDF}
    \end{split}
\end{equation}
where $\Gamma(N)=(N-1)!$ for our positive integer $N$.
This distribution has the mean and covariance which are expected \citep{Kozubowski2013}
\begin{equation}
    \begin{split}
      \text{E}\left[\mathbf{\tilde{M}_{xy}}\right] &=\text{E}\left[\mathbf{\tilde{G}_{xy}}\right] =
      \begin{pmatrix}
          H_r P_s \\ 
          -H_i P_s
      \end{pmatrix} \, \text{, and }\\
      \text{Cov}\left[\mathbf{\tilde{M}}_{xy}\right]   &= \frac{1}{N}\text{Cov}\left[\mathbf{\tilde{G}}_{xy}\right] = 
      \frac{1}{N}
      \begin{pmatrix} 
         \eta + H_r^2P_s^2  &      - H_rH_iP_s^2 \\  
              - H_rH_iP_s^2 & \eta + H_i^2P_s^2  \\
      \end{pmatrix} \,.
    \end{split}
    \label{eqn:n_averaged_exp_cov}
\end{equation}

\subsection{Cospectrum and quadrature spectrum}

We can use the same arguments presented in Section~\ref{sec:co_quad} in order to find the PDFs of the N-averaged cospectrum and quadrature spectrum random variables $\tilde{M}_r$ and $\tilde{M}_i$ \citep{Kozubowski2013},
\begin{equation}
    \begin{split}
        f_{r,N}\left(M_r\right) = \sqrt{\frac{2 }{\pi \eta  }} 
        &\frac{N^{N+\frac{1}{2}} |M_r|^{N-\frac{1}{2}}}{ \Gamma{(N)} } 
        \left({H_r^2 P_s^2+2\eta} \right)^{\frac{1-2N}{4}}
        \text{e}^{ \frac{N P_s H_r}{\eta} M_r } \\
        &\times K_{N-\frac{1}{2}}\left(\frac{N}{\eta }  \sqrt{H_r^2 P_s^2+2 \eta }|M_r|\right) \, ,
    \end{split}
    \label{eqn:mr_pdf}
\end{equation}
and
\begin{equation}
    \begin{split}
        f_{i,N}\left(M_i\right) = \sqrt{\frac{2 }{\pi \eta  }} 
        &\frac{N^{N+\frac{1}{2}} |M_i|^{N-\frac{1}{2}}}{ \Gamma{(N)} } 
        \left({H_i^2 P_s^2+2\eta} \right)^{\frac{1-2N}{4}}
        \text{e}^{ \frac{-N P_s H_i}{\eta} M_i } \\
        &\times K_{N-\frac{1}{2}}\left(\frac{N}{\eta }  \sqrt{H_i^2 P_s^2+2 \eta }|M_i|\right) \, .
    \end{split}
    \label{eqn:mi_pdf}
\end{equation}
The expectation and variances of $\mathbf{\tilde{M}}_r$ and $\mathbf{\tilde{M}}_i$ trivially come from Eq.~\ref{eqn:n_averaged_exp_cov}.

\subsection{Magnitude and Phase Coordinates}

As we did for the unaveraged case, we consider the magnitude and phase of an N-averaged cross spectrum with the random variable $\mathbf{\tilde{M}}_p$. 
Again, we can transform our distribution of $\mathbf{\tilde{M}}_{xy}$ into the distribution of $\mathbf{\tilde{M}}_p$
\begin{equation}
    \begin{split}
        f_{\mathbf{M}_p,N}\left(\rho_M, \Delta_{M\phi}\right) =
        \frac{N^{N+1}\rho_M^{N}}{\pi \eta \Gamma(N) } & \left( |H|^2 P_s^2 + 2\eta\right)^\frac{1-N}{2} \text{e}^{\frac{N P_s}{\eta}|H|\rho_M\cos{\Delta_{M\phi}}}  \\
        &\times K_{N-1}\left(\frac{N \sqrt{|H|^2 P_s^2+2 \eta }}{\eta }\rho_M\right)\,.
    \end{split}    
    \label{eqn:N_mag_arg_av_PDF}
\end{equation}

\subsubsection{Marginalised N-averaged Magnitude}

\begin{figure*}
    \includegraphics[]{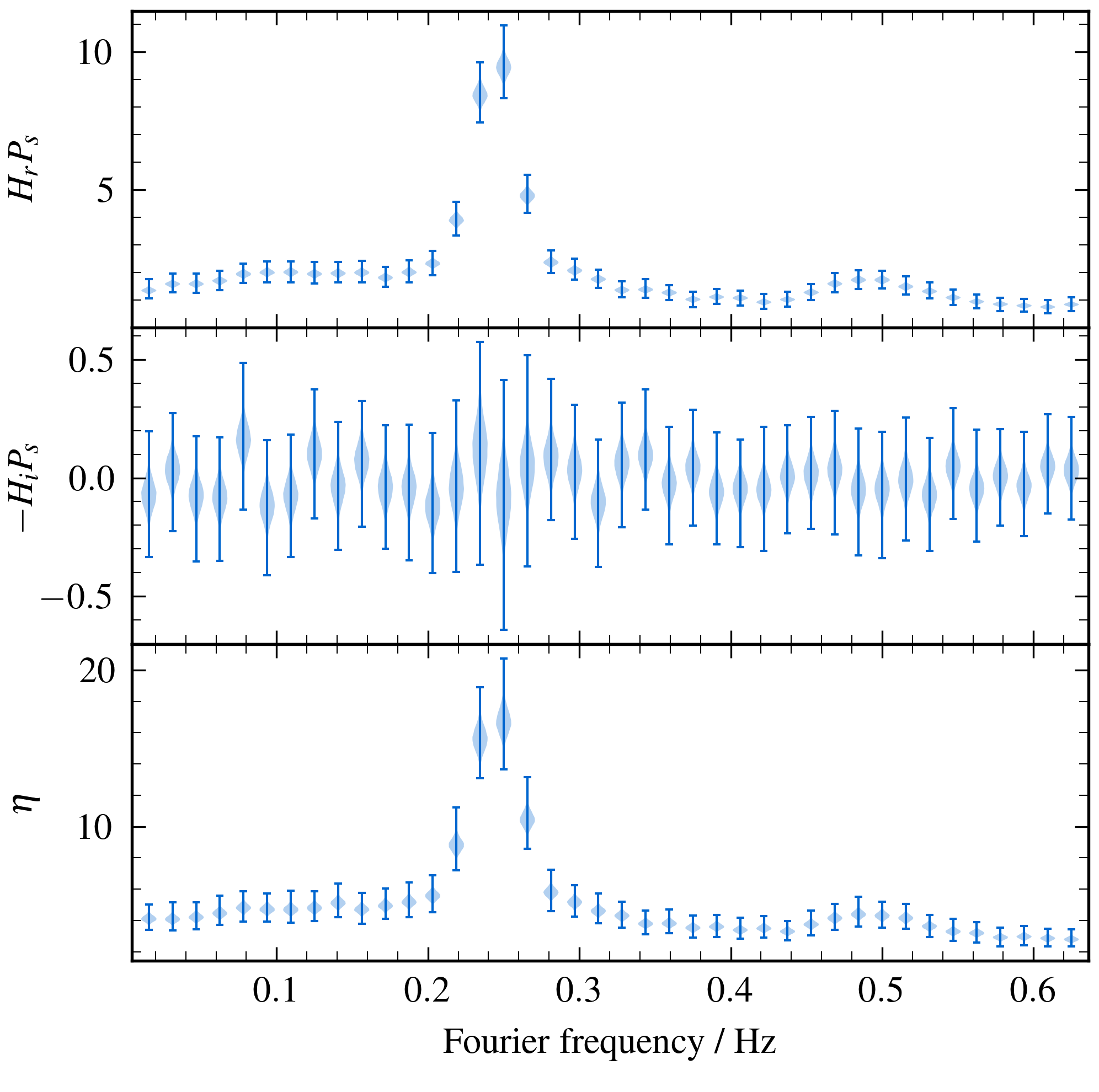}
    \includegraphics[]{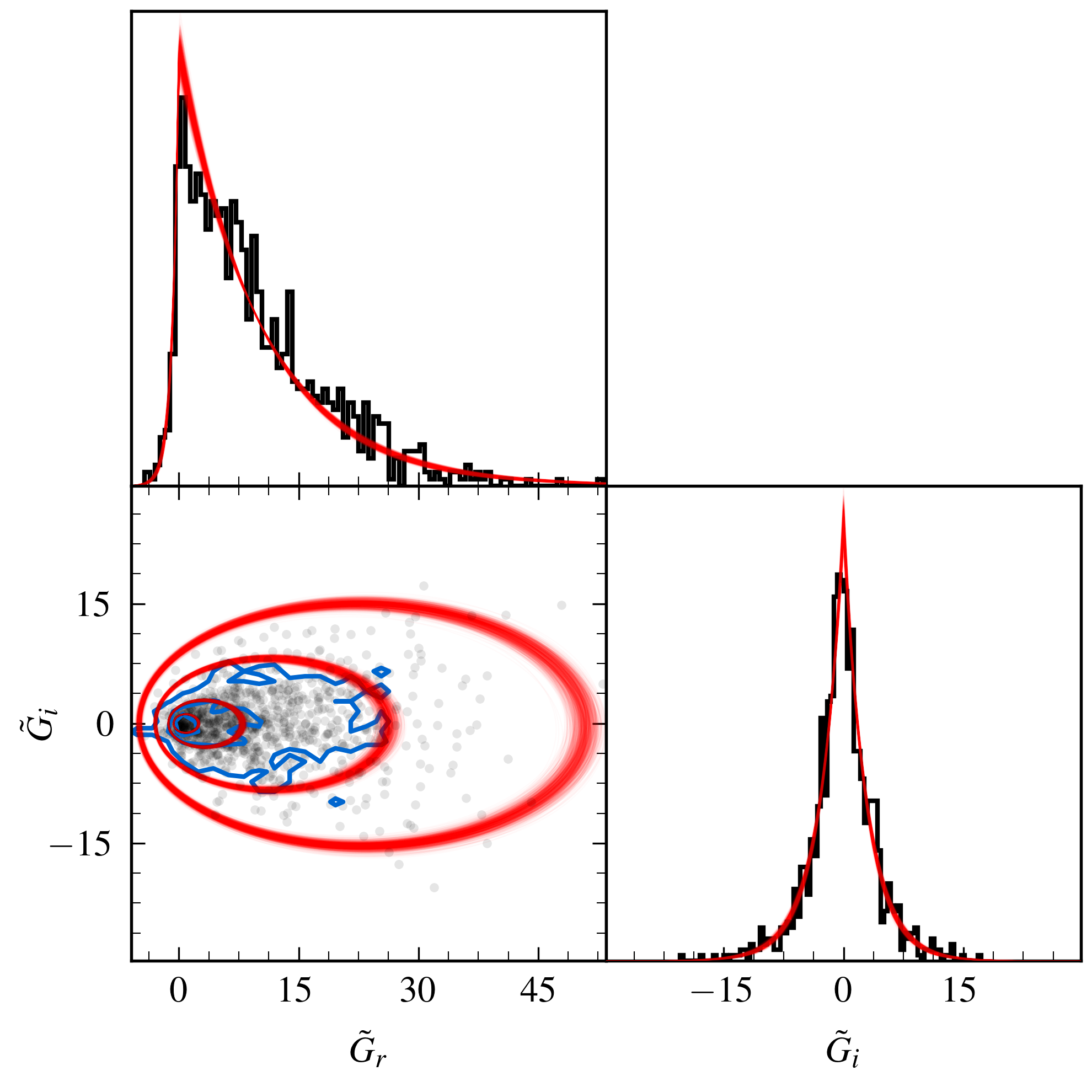}
    \caption{\textbf{Left:} An example cross spectrum, calculated between the light curves from the FPMA and FPMB detectors of a \textit{NuSTAR} observation of H~1743-322 which shows a strong QPO.  For the frequency bins shown, we use MCMC maximum likelihood estimation to fit the parameters $H_rP_s$, $-H_iP_s$, and $\eta$, which correspond to the mean of the co and quadrature spectra, and the spread.  The violin shapes represent the posterior distribution of the parameters sampled at each frequency bin; the `error-bar' shows the full range of the 100,000 MCMC samples used.
    \textbf{Right:} We show the MCMC as fit to the $0.25$~Hz frequency bin (corresponding to the peak of the QPO). The $0.5$, $1$, and $2~\sigma$ contours from the data are shown in blue, while the $0.5$, $1$, $2$, and $3~\sigma$ contours of the distribution are over-plotted in red. We show the contours from many parameter draws from the MCMC, transparently, to show the sampled spread. The PDFs of the real and imaginary parts of the cross spectrum are also over-plotted onto their respective histograms, from the same MCMC samples.}
    \label{fig:mcmc_fit}
\end{figure*}

Marginalising Eq.~\ref{eqn:N_mag_arg_av_PDF} over the angular coordinate $\Delta_\phi$ allows us to find the PDF for the amplitude of the averaged cross spectrum
\begin{equation}
    \begin{split}
        f_{\rho_M,N} \left(\rho_M\right) = \frac{2 N^{N+1} \rho_M^N  }{\eta  \Gamma(N)} &\left(|H|^2  P_s^2+2 \eta \right)^{\frac{1-N}{2}} I_0\left(\frac{N |H| P_s }{\eta }\rho_M\right) \\
        &\times K_{N-1}\left(\frac{N \sqrt{|H|^2 P_s^2+2 \eta }}{\eta }\rho_M\right)  \,.
    \end{split}
    \label{eqn:N_mag_PDF}
\end{equation}
Therefore, the expectation and variance of the random variable $|\tilde{M}|$ are given by
\begin{equation}
    \begin{split}
    \text{E}\left[\tilde\rho_M\right] = &\sqrt{\pi}\eta^N\frac{2^{N-1}\Gamma(N+\frac12)}{\Gamma(N+1)}\left(|H|^2P_s^2+2\eta\right)^{-\left(N+\frac12\right)} \\
    &\times \Bigg(\left(|H|^2P_s^2+2\eta\right) \,{_2F_1}\left[-\frac12,N+\frac12,1,\frac{|H|^2P_s^2}{|H|^2P_s^2+2\eta}\right] \\
    & \quad\, + 2N|H|^2P_s^2 \,{_2F_1}\left[\frac12,N+\frac12,1,\frac{|H|^2P_s^2}{|H|^2P_s^2+2\eta}\right] \Bigg) \,, \\    
    \end{split}
    \label{eqn:mag_M_expectation}
\end{equation}
\begin{equation}
    \text{Var}\left[\tilde\rho_M\right] = \frac{(N+1) |H|^2 P_s^2 + 2\eta}{N} - \left(\text{E}\left[\tilde\rho_M\right]\right)^2 \,, 
    \label{eqn:mag_M_variance}
\end{equation}
where ${_2F_1}$ is the ordinary hypergeometric function \citep[Eqs. \href{https://dlmf.nist.gov/15.1.E1}{15.1.1} and \href{https://dlmf.nist.gov/15.2.E1}{15.2.1};][]{DLMF}.
The magnitude of the averaged cross spectrum is, like its unaveraged counterpart, a biased estimator (see Section~\ref{sec:biased_magnitude} and Appendix~\ref{app:BP10errs}).

\subsubsection{N-averaged Phase}

We are unable to provide a general formula for the marginalised distribution for the phase of the N-averaged cross spectrum. However, we are able to compare our distribution to the \citet{Bendat2010} result for the uncertainty on the phase
\begin{equation}
     \delta \phi \approx \sqrt{\frac{1-g^2}{2g^2 N}}\,,
\end{equation}
where $g^2$ is the raw coherence (see Section~\ref{sec:coherences}). We present this comparison in Appendix~\ref{app:BP10errs}.

\section{Example use cases}
\label{sec:discus}

\subsection{Multiple detectors with instrumental dead time}

Some modern X-ray telescopes, such as \textit{NuSTAR} \citep{Harrison2013} and IXPE \citep{Weisskopf2016} have multiple, co-aligned, independent detectors.  In this case, each detector would record the same underlying signal, but has independent noise terms.  As detectors are not perfect, they can suffer from instrumental dead time: a short period of time after an event in which incoming photons cannot be detected.  In the Fourier domain, this can show up as a `wavey' pattern in the power spectrum \cite{Vikhlinin1994, Zhang1995}.  In the case of \textit{NuSTAR}, which has two detectors FPMA and FPMB, \citet{Bachetti2015} initially suggested the use of the cospectrum (which, as we have discussed, has an expectation value free of Poisson noise) to remove the effects of dead time.  However, they later noted that the cospectrum would still have an RMS modulation due to the dead time \citep[][hereafter \citetalias{Bachetti2018}]{Bachetti2018}.

Following \citetalias{Bachetti2018}, we consider the signal recorded by two detectors as each being convolved with a dead time kernel, as described by \citet{Vikhlinin1994}, which does not depend on the exact realisation of photon arrival times but rather the overall count rate and detector characteristics. In Fourier space, this can be represented as a multiplication by a factor $D$.
In the case of signals from two identical detectors observing the same source, the underlying signal observed is the same. Therefore, the transfer function between the signals is\footnote{If the normalisation of the detectors is slightly different, $H$ becomes a real number related to the difference.} $H=1$. Then, the uncorrelated part of the signal is only the respective photon count noise in the detectors $N_{A}$, $N_{B}$, and the cross spectrum between the signal detectors `A' and `B' therefore becomes
\begin{equation}
    G_\text{deadtime} = |D|^2 G_\text{livetime}\, ,
\end{equation}
where $G_\text{livetime}$ is the cross spectrum that would have been detected if the detectors did not suffer from dead time. 
This leads to a scaling of \textit{both} the noise and signal powers by a factor of $|D|$.
As, in the absence of dead time, the expected power of the photon counting noise is fixed by the photon rate (in any sensible normalisation), the powers $P_{N_A}$ and $P_{N_B}$ are known.  This causes a straight forward effect on the distribution, and the effect of $|D|$ can be extracted when both the real and imaginary parts of the cross-spectrum are considered.  This is, in principle, a very similar idea as the Fourier Amplitude Difference method presented by \citetalias{Bachetti2018}.

\subsection{A case study: H~1743-322}

\begin{figure*}
    \centering
    \includegraphics[]{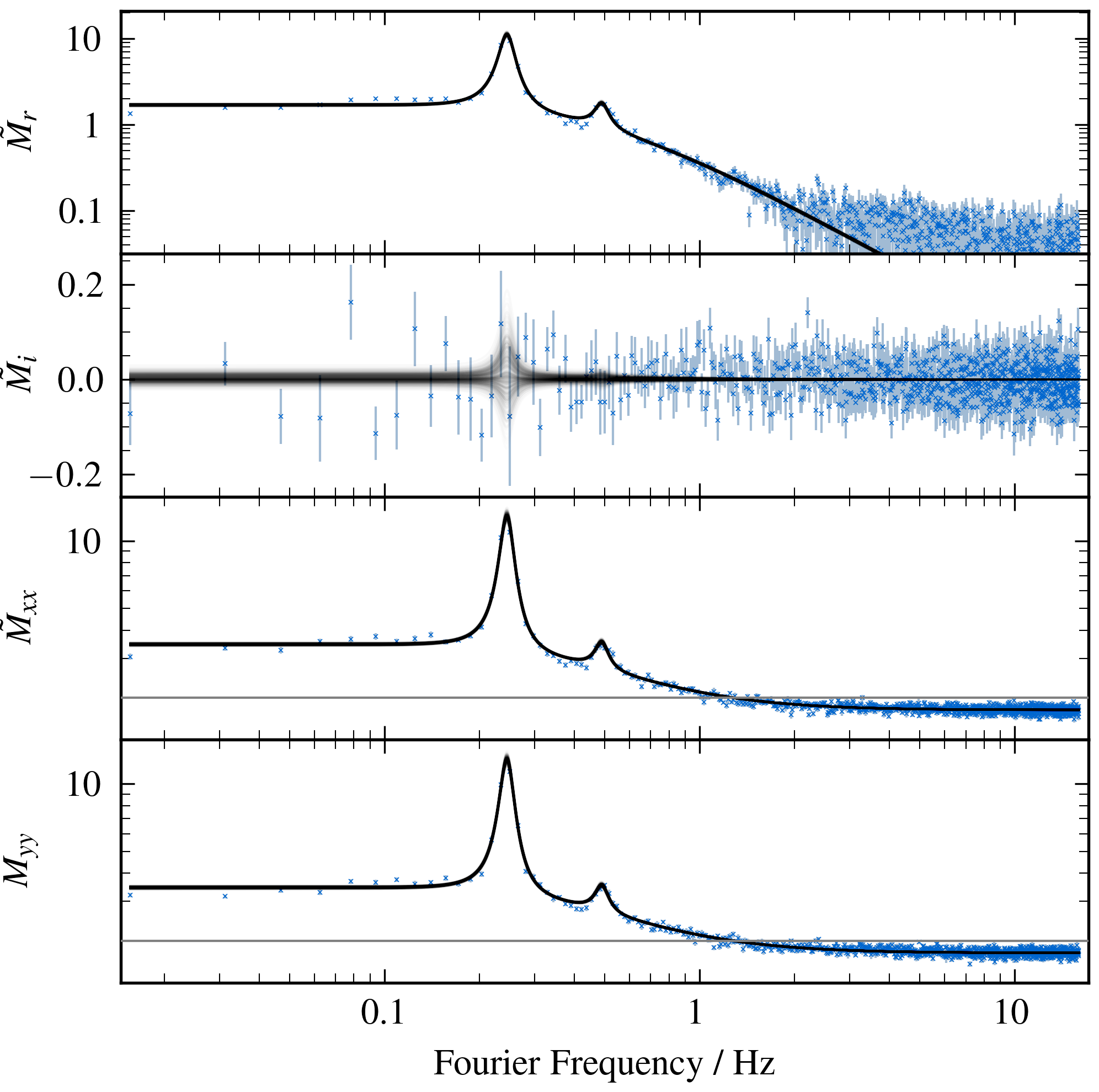}
    \includegraphics[]{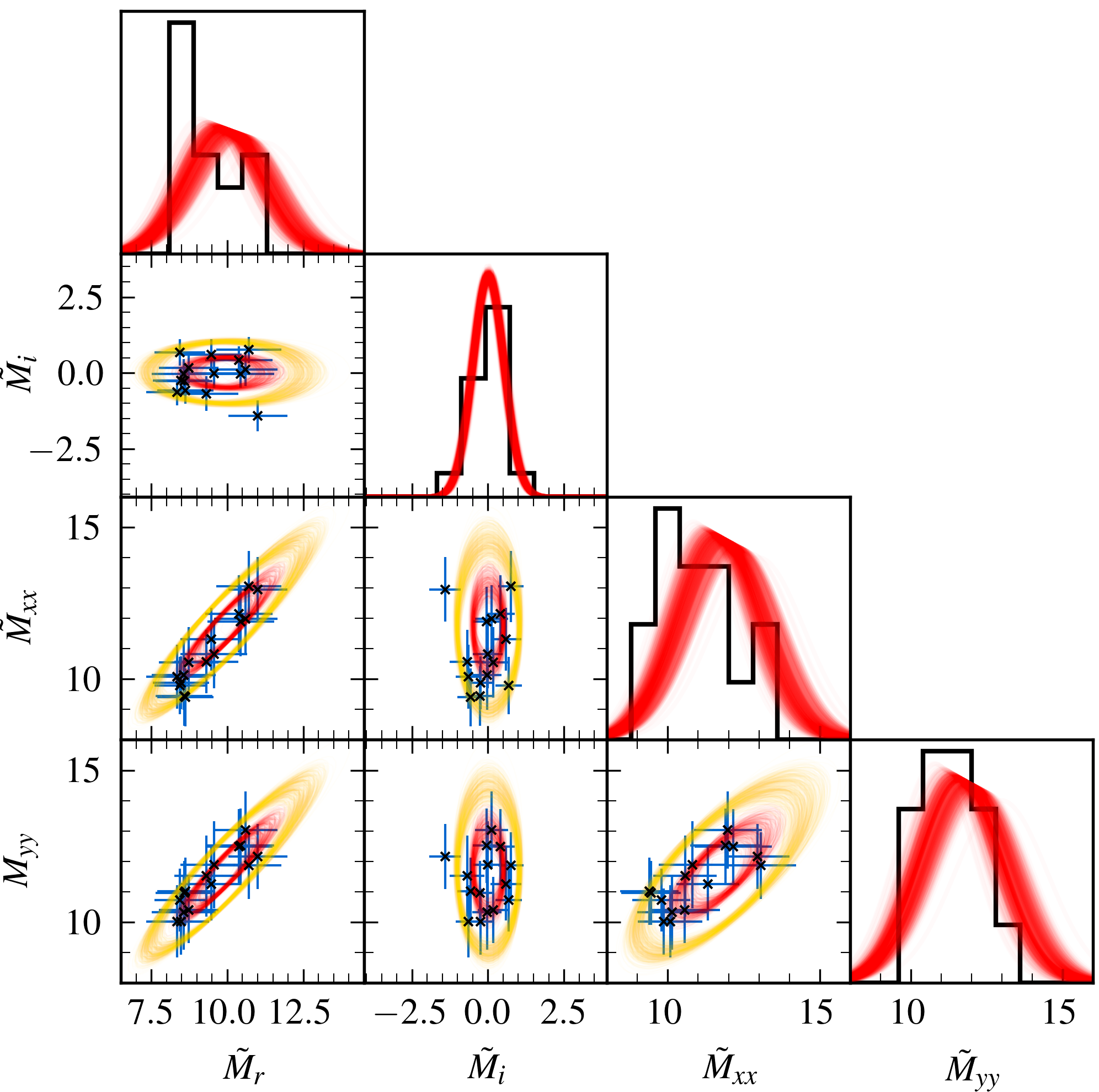}
    \caption{\textbf{Left:} Using the same observation as in Fig.~\ref{fig:mcmc_fit}, when creating the cross spectra we now average spectra created from groups of 73 light curve segments, meaning each frequency has 15 realisations of the $N=73$ cross spectrum.  We use the central limit theorem to fit a model of 3 Lorenztian functions, and an empirical dead time model, to simultaneously fit the power spectra of each light curve and also the co and quadrature spectra. For each frequency, we have 15 realisations of the averaged quantities, the mean and standard deviation of which are plotted with the blue crosses and light-blue error bars.
    We overplot 500 samples of the model from a (much longer) MCMC in translucent black, to show the best fitting model. The expected Poisson noise power in the power spectra is indicated with a horizontal grey line; dead time suppression means that the actual noise is lower than expected.
    \textbf{Right:} We take the 15 realisations of the $N=73$ averaged cross spectra, and take the cospectrum and power spectra from the frequency bin closest to} the QPO frequency.  The markers show the mean spectra of the 15 realisations, and the error bars show the standard error on the mean. We use the model parameters of same 500 samples of the MCMC as shown in (left) to reconstruct the 4D normal distribution at the QPO frequency; which is plotted as the marginal 2D distribution of each pair of variables, with the red and gold contours showing the $1$ and $2\sigma$ levels respectively. We also show the 1D histograms of the points, with the respective marginal distributions.
    \label{fig:averaging_fit}
\end{figure*}

To demonstrate the use of these statistics, we consider a case study of a \textit{NuSTAR} observation of the black hole X-ray binary H~1743-322.  The observation \citep[ObsId: 80001044004;][]{Ingram2016} is known to have a light curve that contains a strong type-C low frequency quasi-periodic oscillation (QPO) with a fundamental frequency of $\sim0.25$~Hz, which sits on top of `broad band noise' which is intrinsic to the signal itself \citep{Ingram2016}.

We will examine two separate cases. To start, we will look at the distribution of $\tilde{G}_{xy}$ at different frequencies around the QPO. This will allow us to extract the the parameters $H_rP_s$, $-H_iP_s$, and $\nu$ for each frequency bin (effectively the means and covariance of the co and quadrature spectra in each frequency bin).  We will then introduce the case where we fit a simple parameterise model to be jointly fit to the power, co, and quadrature spectra, i.e. to $\mathbf{\tilde{G}}(\nu_j)$; to do this we will use averaging over a number of light curve segments such that we can use Gaussian statistics.

\subsubsection{Fitting the unaveraged cross spectrum}

We start by producing the cross spectrum between the light curves of FPMA and FPMB.  For this, we split the light curves in 64~s long segments, made with time bins $1/32$~s wide.  This results in 1095 segments of the light curve (and therefore 1095 realisations of the power and cross spectra for each frequency bin), with a Nyquist frequency of $16$~Hz and minimum frequency resolution of $1/64$~Hz.
Independently for each frequency, we maximise the likelihood with respect to three parameters,  $H_rP_s$ (mean of the cospectrum), $-H_iP_s$ (mean of the quadrature spectrum), and $\eta$.  The total likelihood for each frequency bin is given by
\begin{equation}
    \begin{split}
        \mathcal{L}_\text{total}\big(H_rP_s, -H_iP_s, \eta ; \{G_r,G_i\}\big) &= \prod_{n=1}^{1095} \mathcal{L}\big(H_rP_s, -H_iP_s, \eta ; \tilde{G}_{r,n},\tilde{G}_{i,n}\big) \\
        &= \prod_{n=1}^{1095} f\big(\tilde{G}_{r,n},\tilde{G}_{i,n}; H_rP_s, -H_iP_s, \eta\big) \, ,
     \end{split}
\end{equation}
where $f\big(G_{r,n},G_{i,n}; H_rP_s, -H_iP_s, \eta\big)$ comes directly from the PDF Eq.~\ref{eqn:PDF}. 
For computational reasons, in practice we computed and summed the log-likelihoods.

We use Markov chain Monte Carlo (MCMC) to explore the likelihood space and constrain the three parameters.  In Fig.~\ref{fig:mcmc_fit} (left) we show the MCMC parameters for a number of frequency bins around the QPO frequency.  
We clearly see the significant increase in coherent signal power at the QPO frequency ($\sim2.5$~Hz), which also causes an increase in the value of $\eta$. These increases also show as increases in the distribution widths, most notable for the quadrature spectrum ($-H_i P_s$) which has a mean value of $0$ due to the simultaneous observations of \textit{NuSTAR}'s FPMA and FPMB detectors, resulting in no phase lag.  However, despite the expected value of 0, the variance in the quadrature spectrum is still increased around the QPO frequency (c.f. Eq.~\ref{eqn:var_GrGi}).  We highlight how the distribution has been estimated for the $0.25$~Hz frequency bin, which contains the peak of the QPO, in Fig.~\ref{fig:mcmc_fit} (right). We show the 2D histogram of the real and imaginary parts of the 1095 samples of the cross spectrum at this frequency, compared to the distribution drawn from many different samples of the MCMC.
We see that the MCMC simulation fits our analytical distributions (red) to the distribution of the cross spectrum realisations calculated from the available light curve segments 
(shown as the black stepped lines, black dots, and blue contours). 
We also see that the analytical estimate of the distribution varies only a small amount with the spread of values in the MCMC (i.e. the red lines are all similar to one another), meaning that it is appropriate to empirically estimate the distributions from an MCMC simulation that uses all segments, and uncertainties on the parameters derived from the MCMC (shown in Fig.~\ref{fig:mcmc_fit} (left)) are sensible.

\subsubsection{Including the power spectra by averaging}
\label{sec:h1743ave}

To make use of the power spectra alongside the cross spectrum, we have to employ Gaussian statistics since we were unable to derive a general joint distribution for the cross spectrum and the two power spectra. Therefore, we are required to sufficiently average our spectra such that the central limit theorem is a valid approximation.
For the same H~1743-322 example, we compute 15 sets of averaged cross and power spectra, with each averaged of 73 realisations (i.e. $15\times73=1095$), such that the cross spectrum of the $j^\text{th}$ frequency bin is $\mathbf{\tilde{M}}_{j,k}=\frac{1}{73}\sum_{n=1}^{73} \mathbf{\tilde{G}}_{j,n}$.  
Here, the $k^\text{th}$ index represents one of 15 realisations of the averaged cross spectrum, where the order of the $1095$ realisations of $\mathbf{\tilde{G}}_{j,n}$ have been randomised before the averaging (such that each of the 15 realisations of $\mathbf{\tilde{M}}_{j,k}$ has not been averaged from sequential realisations of $\mathbf{\tilde{G}}_{j,n}$).

We fit a model comprised of three Lorenztians, one to account for the broadband noise, and one each for the two QPO harmonics; a multiplicative `sinc' function for the effects of dead time \citep{Bult2017}; and the coherence and phase (parametrised by the real and imaginary parts of the transfer function $H_r$ and $H_i$) of the overall cross spectrum which is taken to be the same across all frequencies.  Here, the phase difference is expected to be $0$, and the magnitude of $H$ should be directly related to the count rates of FPMA and FPMB.  We used an MCMC simulation to explore the likelihood in the Gaussian limit. The total likelihood comes from the product of the likelihoods at each frequency
\begin{equation}
    \mathcal{L}_\text{total}(\theta) = \prod_j \prod_{k=1}^{15} \mathcal{L}_j\big(P_{sj}, P_{u_xj},P_{u_yj}, H_j; \mathbf{\tilde{M}}_{j,k} \big) \, .
\end{equation}
Here, $\mathcal{L}$ refers to the likelihood from one frequency bin (see Eq.~\ref{eqn:averaged_likelihood}), and the parameters at each frequency bin $\left\{P_{sj}, P_{u_xj},P_{u_yj}, H_j\right\}$) are calculated from the model parameters $\theta$. 
Using this we are able to constrain the parameters of our model by simultaneously fitting these cross spectra and power spectra as shown in Fig.~\ref{fig:averaging_fit} (left).  In Fig.~\ref{fig:averaging_fit} (right) we show the distribution found at the QPO frequency, although we reiterate that the maximum likelihood fitting was for the model on the whole frequency range.

\section{Conclusions}

In this paper, we used the fact that the power spectra and cross spectra can be written as quadratic forms of normal random variables in order to find the statistical distribution of the cross spectrum, which is currently missing from the astrophysics literature. In general, we considered the statistical properties when considering only one frequency bin, however due to the statistical independence of each frequency bin, this can be easily applied to the spectrum across the entire frequency range (with the exception at the Nyquist frequency, and at $0~Hz$, which will have different statistical distributions than derived here). Throughout this paper, we make three key assumptions: the first is typically assumed within X-ray timing work, that that each light curve is stationary, such that the underlying properties are unchanging. The second assumption is that there is only a linear correlation between the two light curves. The final assumption is that the two light curves are entirely independent, for example coming from two independent detectors; while selecting two energy bands from the same light curve will select different photons, beware of detector characteristics which may mean these light curves may not be entirely independent.

To investigate the statistics underlying the cross spectrum, we started by developing constructions for the means, and covariance matrix, for the 4D variable containing the power spectra of each signal, and the cospectrum and quadrature spectra (i.e. the real and imaginary parts of the cross spectrum) $\mathbf{\tilde{G}}=\left(\tilde{G}_{xx},\tilde{G}_{yy},\tilde{G}_r,\tilde{G}_i\right)'$, recovering the results of \citet{Bendat2010}. In the case of averaging enough realisations, such as binning between frequencies or between the cross spectrum calculated from different segments of the light curve, we note that multi-variate Gaussian statistics can be used due to the central limit theorem. However, we highlight that there is intrinsic covariance between the power, co, and quadrature spectra for each frequency bin that must be included in this calculation.  While this is not an explicitly new result, the covariance between the co and quadrature spectra has typically been incorrectly ignored when simultaneously fitting \citep[e.g. ][]{Mendez2024, Konig2024}, which is particularly problematic when attempting to use $\chi^2$ statistics to claim statistical confidence in a model fit. Only in the case where there is both zero coherence between the two light curves \textit{and} sufficient averaging such that the distributions can be considered Gaussian can this approach be used. We remind the reader that even in the case where the coherence is 0, without averaging $\tilde{G}_r$ and $\tilde{G}_i$ are not independent and follow a 2D Laplace distribution \citepalias[][]{Huppenkothen2018}.

To consider the underlying distribution of a cross spectrum with no averaging or binning, we used the CF. We provided the PDF for the 2D random variable $\mathbf{\tilde{G}}_{xy}=\left(\tilde{G}_r,\tilde{G}_i\right)'$ (Eq.~\ref{eqn:PDF}), and provided the expectation and covariance matrix of $\mathbf{\tilde{G}}_{xy}$. We also showed the marginal distributions of $\tilde{G}_r$ and $\tilde{G}_i$ (Eq.~\ref{eqn:real_imag_PDFs}). As it is often physically interesting to consider the magnitude and phase lag of the cross spectrum, we also provided the distributions for the polar form of the cross spectrum $\mathbf{\tilde{G}}_p=\left(\tilde{\rho}_G,\tilde{\Delta}_\phi\right)'$ (Eq.~\ref{eqn:polar_PDF}). While we also provided the marginal distributions of $\tilde{\rho}_G$ and $\tilde{\Delta}_\phi$ (Eqns.~\ref{eqn:mag_PDF} and \ref{eqn:ang_PDF}), we caution that $\tilde{\rho}_G$ itself is a biased estimator of the underlying distribution. 

The final distributions we consider are those where a small number of realisations of the cross spectrum have been averaged, but not enough to be able to use Gaussian statistics; these could be from neighbouring frequency bins, or from different segments of the light curve. It is important to remember that the assumption is that the underlying distribution of the bins or segments which are averaged is the same. We consider the variable $\mathbf{\tilde{M}}_{xy}=\frac1N\sum_n\tilde{G}_{xy,n}=\left(\tilde{M}_r,\tilde{M}_i\right)'$, and produce the 2D PDF (Eq.~\ref{eqn:N_av_PDF}), and the marginal PDFs of $\tilde{M}_r$ and $\tilde{M}_i$ (Eqs.~\ref{eqn:mr_pdf} and \ref{eqn:mi_pdf}). Finally, in the same manner, we consider the polar form of the low-N averaged cross spectrum $\mathbf{\tilde{M}}_p=\left(\tilde{\rho}_M,\tilde{\Delta}_{M\phi}\right)'$ (Eqs.~\ref{eqn:N_mag_arg_av_PDF} and \ref{eqn:N_mag_PDF}). Again, we sound caution that $\tilde{\rho}_M$ proves to be a biased estimator. We were unable to find the marginal distribution of $\tilde{\Delta}_{M\phi}$, however we show how the 2D distribution compares to the \citet{Bendat2010} results.

We finished by discussing using the cross spectrum to account for dead time effects when using light curves from identical detectors.  We showed a case study to study a \textit{NuSTAR} observation of the black hole X-ray binary H~1743-322, in two ways. To start, we showed how using the unaveraged distributions of $\mathbf{\tilde{G}}_{xy}$ for individual frequency bin can fit the co and quadrature spectra, and we also highlight how the covariance between them is different in different frequency bins, given the underlying signal power at each frequency. We then introduced averaging over subsets of segments of the light curves, such that we can use the approximate Guassian statistics to jointly fit a model to the power spectra and the complex cross spectrum.

We encourage others to make use of these statistics when analysing cross spectra, particularly in the field of X-ray astronomy. Currently, it is common to rely on the sum of computing the sum of $\chi^2$ statistics for the real and imaginary parts of the cross spectrum, which is only appropriate in the case of zero covariance between the real and imaginary parts. We have noted that in general there is an intrinsic non-zero covariance, and thus jointly fitting to real and imaginary parts of the cross spectrum required accounting for the covariance. We provide the equations necessary to do that in the limits of a single-realisation, a cross spectrum averaged over many realisations (such that multivariate Gaussian statistics can be assumed), and averaging over an intermediate number of realisations (such that Gaussian statistics cannot be assumed).

\section*{Acknowledgements}

EN thanks Guglielmo Mastroserio and Michiel van der Klis for valuable discussion.
EN’s research was supported by an appointment to the NASA Postdoctoral Program at the NASA Goddard Space Flight Center, administered by Oak Ridge Associated Universities under contract with NASA. AI acknowledges support from the Royal Society.
The authors thank the anonymous referees, whose comments have helped improve the clarity of the paper.

\section*{Conflicts of Interest}

The authors declare no conflict of interest.

\section*{Data Availability}

The observational data used in this research are public and available for download from the HEASARC.


\section*{Additional Material}
We provide an online repository including implementations of many of the distributions in this paper, and example uses of them. This repository can be found at 
\href{https://github.com/EdNathan/cross-spectrum-statistics-example}{https://github.com/EdNathan/cross-spectrum-statistics-example}.



\bibliographystyle{mnras}
\bibliography{references} 



\appendix

\section{Simulation Procedure}
\label{app:simulations}
Throughout this paper, we use simulations of the cross spectrum to compare to our distributions.  These simulations are done through a similar procedure as described by \citet{Timmer1995}, which we outline here.

The simulations are done by selecting the intended power in each light curve $P_X$ and $P_Y$, which includes the associated amount of Poisson noise $P_{n_x}$ and $P_{n_y}$ (typically we use a noise power of $2$, associated with Poisson noise for power spectra computed with \citet{Leahy1983} normalisation). We also choose a value for the intrinsic coherence of the signal, by selecting the squared coherence $\gamma^2$ and phase difference $\phi$ such that $\gamma=\sqrt{\gamma^2}\text{e}^{\I{}\phi}$.
With these values, we find our distribution values with Eq.~\ref{eqn:notation_transform}. Finally, we also select the number of samples for averaging $N$ (whereby $N=1$ is the unaveraged case).

For each realisation we wish to simulate, we produce $N$ values for the Fourier transform of the underlying signals by drawing normal random variables with 0 mean and a variance of half the required power as so
\begin{alignat}{2}
    \tilde{S}_{n} &= \tilde{S}_{r,n} + \I{} \tilde{S}_{i,n}\,,       \quad &&\tilde{S}_{r,n}\,,\tilde{S}_{i,n}\sim\mathcal{N}\left[0,\frac{1}{2}P_s\right]\,, \notag\\
    \tilde{U}_{x,n} &= \tilde{U}_{x,r,n} + \I{} \tilde{U}_{x,i,n}\,, \quad &&\tilde{U}_{x,r,n}\,,\tilde{U}_{x,i,n}\sim\mathcal{N}\left[0,\frac{1}{2}P_{u_x}\right]\,,\\
    \tilde{U}_{y,n} &= \tilde{U}_{y,r,n} + \I{} \tilde{U}_{y,i,n}\,, \quad &&\tilde{U}_{y,r,n}\,,\tilde{U}_{y,i,n}\sim\mathcal{N}\left[0,\frac{1}{2}P_{u_y}\right]\,. \notag
\end{alignat}
From these random variables, we construct the simulated Fourier transforms
\begin{equation}
    \begin{split}
        \tilde{\mathcal{F}}_{x,n} &=  \tilde{S}_n+\tilde{U}_{x,n} \,,\\
        \tilde{\mathcal{F}}_{y,n} &=  H\tilde{S}_n+\tilde{U}_{y,n} \,,
    \end{split}
\end{equation}
which then allow the creation of the simulated cross spectrum via
\begin{equation}
    \begin{split}
        \tilde{G}_{xx} &=  \frac{1}{N}\sum_n \tilde{\mathcal{F}}_{x,n}\tilde{\mathcal{F}}_{x,n}^* \,,\\
        \tilde{G}_{yy} &=  \frac{1}{N}\sum_n \tilde{\mathcal{F}}_{y,n}\tilde{\mathcal{F}}_{y,n}^* \,,\\
        \tilde{G}_r &=  \frac{1}{N}\sum_n \Re\left[ \tilde{\mathcal{F}}_{x,n}\tilde{\mathcal{F}}_{y,n}^*\right] \,,\\
        \tilde{G}_i &=  \frac{1}{N}\sum_n \Im\left[ \tilde{\mathcal{F}}_{x,n}\tilde{\mathcal{F}}_{y,n}^*\right] \,,\\
        \tilde\rho_G &=  \frac{1}{N}\left|\sum_n \tilde{\mathcal{F}}_{x,n}\tilde{\mathcal{F}}_{y,n}^*\right|\,,\\
        \tilde{\phi}_G &=  \arg\left[\sum_n \tilde{\mathcal{F}}_{x,n}\tilde{\mathcal{F}}_{y,n}^*\right]\,.\\
    \end{split}
\end{equation}
In general, we use the symbol `$\tilde{G}_\bullet$' for quanties relating to an unaveraged cross spectrum (with $N=1$), while for an averaged cross spectrum with $N>1$ use use the symbol `$\tilde{M}_\bullet$'.

In all of our simulations, we simulate many realisations as above with the same underlying distribution. This would be the same as taking the cross spectrum from the same frequency bin calculated from many light curve segments, or taking many bins from a completely flat, `white noise' cross spectrum. 

\section{Mixing matrices}
\label{app:mix_mat}

Here we list the mixing matrices used in the body of the paper.
\begin{align}
    \mathbf{\Gamma}_r = \frac{1}{2}
    &\begin{pmatrix}
        2H_r &    0 &  H_r &  H_i & 1 & 0 \\
           0 & 2H_r & -H_i &  H_r & 0 & 1 \\
         H_r & -H_i &    0 &    0 & 1 & 0 \\
         H_i &  H_r &    0 &    0 & 0 & 1 \\
           1 &    0 &    1 &    0 & 0 & 0 \\
           0 &    1 &    0 &    1 & 0 & 0
    \end{pmatrix} 
    \label{eqn:mix_mat_Gr} \\[0.5ex]
    \mathbf{\Gamma}_i = \frac{1}{2}
    &\begin{pmatrix}
        -2H_i &     0 &  -H_i &  H_r & 0 & -1 \\
            0 & -2H_i &  -H_r & -H_i & 1 &  0 \\
         -H_r &  -H_r &     0 &    0 & 0 & -1 \\
          H_r &  -H_i &     0 &    0 & 1 &  0 \\
            0 &     1 &     0 &    1 & 0 &  0 \\
           -1 &    0 &    -1 &    0 & 0 &   0
    \end{pmatrix} 
    \label{eqn:mix_mat_Gi} \\[0.5ex]
    \mathbf{\Gamma}_X = \hphantom{\frac{1}{2}}
    &\begin{pmatrix}
        1 & 0 & 1 & 0 & 0 & 0 \\
        0 & 1 & 0 & 1 & 0 & 0 \\
        1 & 0 & 1 & 0 & 0 & 0 \\
        0 & 1 & 0 & 1 & 0 & 0 \\
        0 & 0 & 0 & 0 & 0 & 0 \\
        0 & 0 & 0 & 0 & 0 & 0
    \end{pmatrix} 
    \label{eqn:mix_mat_PX} \\[0.5ex]
    \mathbf{\Gamma}_Y = \hphantom{\frac{1}{2}}
    &\begin{pmatrix}
        |H|^2 &     0 &  0 & 0 &  H_r & H_i \\
            0 & |H|^2 &  0 & 0 & -H_i & H_r \\
            0 &     0 &  0 & 0 &    0 &   0 \\
            0 &     0 &  0 & 0 &    0 &   0 \\
          H_r &  -H_i &  0 & 0 &    1 &   0 \\
          H_i &   H_r &  0 & 0 &    0 &   1
    \end{pmatrix}
    \label{eqn:mix_mat_PY}
\end{align}

\section{Marginalising a characteristic function}
\label{app:marginalising_char}
The definition of the CF for a bivariate PDF $f(z_1, z_2)$ is
\begin{equation}
        \varphi(t_1, t_2) = \iint_{\Re^2} \text{e}^{\I{} (t_1 z_1 + t_2 z_2)} f(z_1, z_2) \dd{z_1}\dd{z_2}  .
\end{equation}
Setting $t_2=0$, the CF $\varphi_1(t_1)$ of the marginalised PDF is $f_1(t_1)=\int_{-\infty}^{\infty} f(z_1, z_2)  \dd{z_2}$, which is demonstrated with
\begin{equation}
    \begin{split}
        \varphi(t_1, 0) &= \iint_{\Re^2} \text{e}^{\I{} (t_1 z_1)} f(z_1, z_2) \dd{z_1}\dd{z_2} \\
        &= \int_{-\infty}^{\infty} \left[\int_{-\infty}^{\infty} f(z_1, z_2)  \dd{z_2}  \right]  \text{e}^{\I{} (t_1 z_1)} \dd{z_1} \\
        &= \int_{-\infty}^{\infty} f_1(z_1)  \text{e}^{\I{} (t_1 z_1)} \dd{z_1} \\
        &= \varphi_1(t_1) \, .
    \end{split}
\end{equation}

\section{Comparison of Bendat and Piersol errors}
\label{app:BP10errs}
\begin{figure*}
    \includegraphics[]{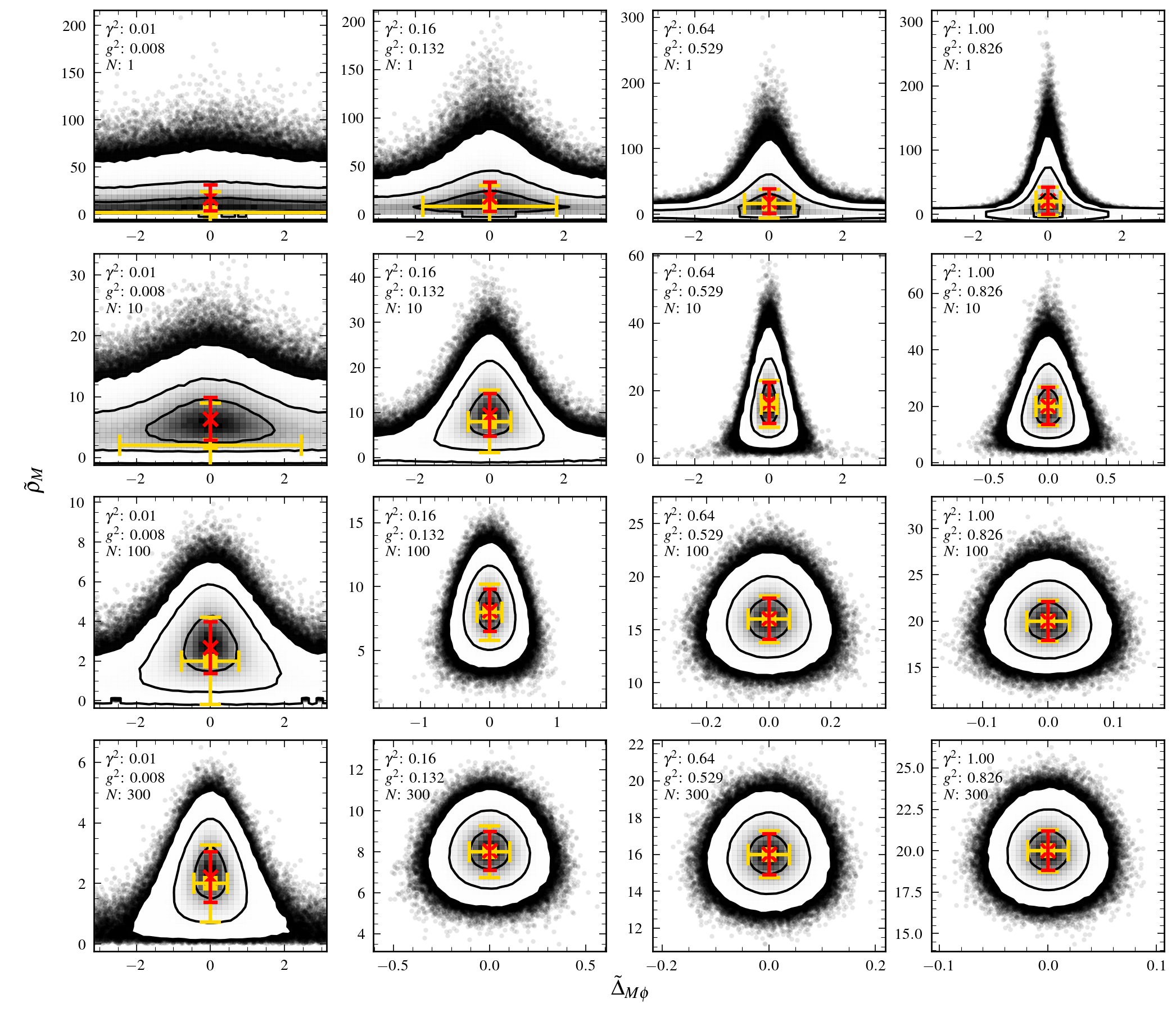}
    \caption{Simulation of $1,000,000$ cross spectral bins, with the simulation procedure in Appendix~\ref{app:simulations}. The underlying powers are chosen as $P_X=P_Y=22$, including $P_{n_x}=P_{n_y}=2$. Four different intrinsic squared coherences are chosen (as labelled, with the resultant raw squared coherence also given), with four values for the averaging (including the unaveraged case $N=1$). The 1, 2, and 3 $\sigma$ contours are shown in each case. For each distribution, we show the \citet{Bendat2010} errors in gold, with our error in red (see text for details).}
    \label{fig:bp10_err_comparison}
\end{figure*}

While we have not provided formulae for the variance of the cross spectral phase,  we can compare them to simulations. In Fig.~\ref{fig:bp10_err_comparison} we show simulations for varying values of coherence and averaging numbers. We also show comparisons of our formulae to those of \citet{Bendat2010}.

\citet{Bendat2010} presented approximate formulae for the random errors on an estimate of the amplitude and phase of a highly averaged cross spectrum as
\begin{equation}
    \begin{split}
    \delta\tilde\phi &= \sqrt{\frac{1-g^2}{2g N}}\,, \\
    \delta|\tilde{M}| &= \frac{|M|}{\sqrt{g^2 N}}\,.
    \end{split}
    \label{eqn:bp10errs}
\end{equation}

However, they only provide the expectation for a measurement of the magnitude squared, i.e. $\text{E}\left[|\tilde{M}|^2\right]$. The gold, square markers in Fig~\ref{fig:bp10_err_comparison} show the underlying value of $|M|$, i.e. what an unbiased estimate should provide; the gold error bars are from Eq.~\ref{eqn:bp10errs} using the known values of $g$ and $|M|$. The red, X markers show the expectation of the distribution of $|\tilde{M}|$, along with the standard deviation on the measurement, from Eqs.~\ref{eqn:mag_M_expectation} and \ref{eqn:mag_M_variance}.

\section{Cross spectral modelling}
\label{app:cross_spec_mod}

\begin{figure*}
    \centering
    \includegraphics[]{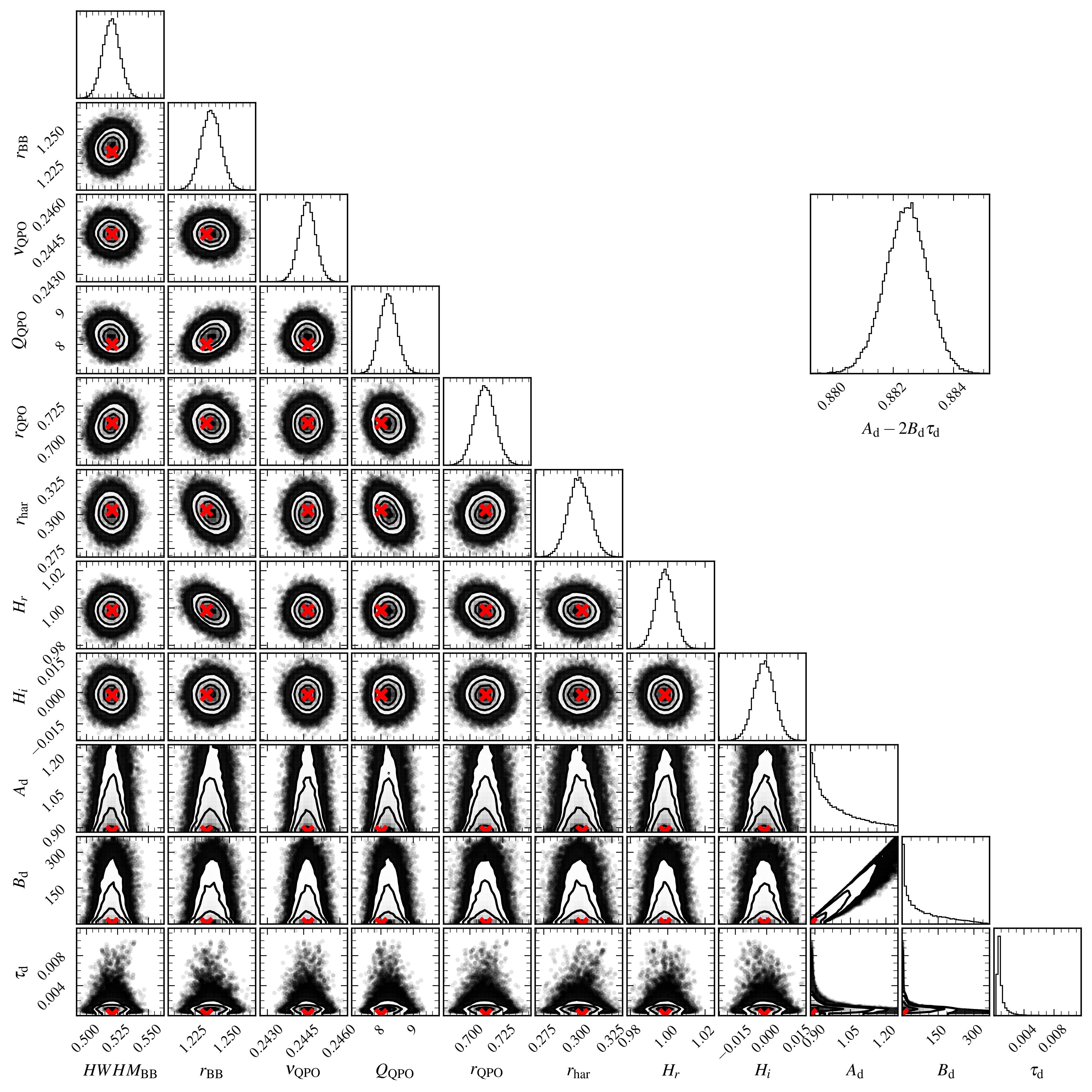}
    \caption{We present samples from the MCMC for each of the parameters of the model which we fit to the cross spectrum, see the text for details of this model. We also reconstruct the combined distribution of $A_d-2B_d\tau_d$, which we argue is the only constrained element of the deadtime model due to the Nyquist frequency of the cross spectrum that we use. For each distribution, the 1, 2, and 3 $\sigma$ contours are shown. We also show the value of the best fitting parameters with the red crosses, found through maximum likelihood estimation.}
    \label{fig:all_param_corner}
\end{figure*}

The model use to fit the averaged cross spectrum in Section~\ref{sec:h1743ave} is built of the sum of three Lorentz functions: one centred at 0~Hz, to fit the broad band noise, with the half width at half maximum $HWHM_\textrm{BB}$ and RMS $r_\textrm{BB}$ as free parameters; one to fit the QPO fundamental harmonic, with central frequency $\nu_\textrm{QPO}$, quality factor $Q_\textrm{QPO}$, and RMS $r_\textrm{QPO}$ as free parameters; and one to fit the QPO second harmonic with only the RMS $r_\textrm{har}$ as a free parameter ($\nu_\textrm{har}=2\nu_\textrm{QPO}$, $Q_\textrm{har}=Q_\textrm{QPO}$).
The sum of these three Lorentz functions define the signal power in each frequency bin, and is therefore 
\begin{equation}
\begin{split}
    P_{sj}&=P_s(\nu_j)\\
    &=L_0(\nu_j;HWHM_\textrm{BB},r_\textrm{BB}) + L(\nu_j;\nu_\text{QPO},Q_\textrm{QPO},r_\textrm{QPO}) \\
    &\qquad+ L(\nu_j;2\nu_\text{QPO},Q_\textrm{QPO},r_\textrm{har})\,
\end{split}
\end{equation}
where $L(\nu;\nu_0,Q,r)$ is a Lorentz function centred at $\nu_0$, with quality factor $Q$ and a total RMS $r$, and $L_0(\nu;HWHM,r)$ is instead a Lorentz function centred at 0~Hz with a given HWHM. 
As the data comes from two identical detectors, and we work in the \citet{Leahy1983} normalisation, the uncorrelated element of each signal is only the Poisson noise $P_{u_xj}=P_{u_yj}=2$.  As a simple model, we set the transfer function to be constant in frequency, with $H_j=H_{r,j}+\I{}H_{i,j}=H_r+\I{}H_i$ where $H_r$ and $H_i$ are free parameters.  Finally, we include the deadtime based upon a $\sinc$ function \citep{Bult2017}
\begin{equation}
    D_j=D(\nu_j;A_d,B_d,\tau_d) = A_d-2B_d\tau_d \sinc{\left(2\pi\nu_j\tau_d\right)}\,.
\end{equation}

We construct the likelihood with Eq.~\ref{eqn:averaged_likelihood}, based upon our 11 model parameters, where
\begin{equation}
    \begin{split}
        \mathbf{m}_{\theta j} &= D_j
          \begin{pmatrix}
            P_{sj} + 2 \\ (H_r^2+H_i^2) P_{sj} +2 \\ H_r P_{sj} \\ -H_i P_{sj}
          \end{pmatrix} \, \text{, and }\\
        \mathbf\Lambda_{\theta_j} &=  D_j^2
          \begin{pmatrix}
               0 & -2\eta &    0 &    0 \\ 
          -2\eta &      0 &    0 &    0 \\
               0 &      0 & \eta &    0 \\ 
               0 &      0 &    0 & \eta \\
          \end{pmatrix} \,.
    \end{split}
\end{equation}

When fitting the data, we first maximised the log-likelihood $\ln\mathcal{L}$, before using the log likelihood to run an MCMC.  Using a prior, all parameters (except $H_r$ and $H_i$) were restricted to be $>0$. We also used top-hat priors for the deadtime parameters $0.5<A_d<1.25$, and $0.0005<\tau_d<0.025$~s.

Our MCMC employed the \citet{Goodman2010} `stretch-move' algorithm \citep{Foreman-Mackey2013}, with 500 walkers ran for 20,000 steps. The main result was demonstrated in Fig.~\ref{fig:averaging_fit}, which shows the last step of each walker. We also present the parameter space as explored by the MCMC in Fig.~\ref{fig:all_param_corner} (taking the final 100 samples from each walker after thinning by a factor of 100). 
The the maximum likelihood values are also shown as red crosses.  

As the main expected deadtime effects occur at higher frequencies than our Nyquist frequency of 16~Hz, the main effect of the deadtime is an overall reduction of power by a factor $\sim (A_d-2B_d\tau_d)$. We therefore include the histogram of $A_d-2B_d\tau_d$ calculated from our chain, which is estimated as $0.8824\pm7\times10^{-4}$.
This is similar to (albeit not consistent with) the rough approximation of using the \texttt{deadc} deadtime estimate $\sim0.847$ provided by the \textit{NuSTAR} pipeline for each light curve. We also have the estimate $\tau_d=0.7_{-0.2}^{+0.6}$~ms, which smaller than the expected typical instrumental deadtime of \textit{NuSTAR} \citep[$\sim2.5$~ms, ][]{Bachetti2015}; as discussed, we do not expect this to be a good constraint because the Nyquist frequency of the cross spectrum we use is below that of the frequency of the typical deadtime (i.e. our sampling rate $\Delta t = 1/32\approx31$~ms is greater that the typical instrumental deadtime $\sim2.5$~ms).


\bsp	
\label{lastpage}
\end{document}